\documentclass[fleqn,usenatbib]{mnras}

\usepackage{newtxtext,newtxmath}

\usepackage[T1]{fontenc}

\DeclareRobustCommand{\VAN}[3]{#2}
\let\VANthebibliography\thebibliography
\def\thebibliography{\DeclareRobustCommand{\VAN}[3]{##3}\VANthebibliography}

\usepackage{graphicx}	
\usepackage{amsmath}	
\usepackage{lscape}
\usepackage{rotating}
\usepackage{float}
\hypersetup{colorlinks = true, linkcolor = blue, urlcolor=blue, citecolor=blue} 
\usepackage[normalem]{ulem}
\usepackage{tabularx}
\usepackage{rotating, booktabs}
\usepackage{inputenc}
\usepackage{geometry}
\usepackage{changepage}
\usepackage{color}

\def\msun{\hbox{M$_\odot$}}
\def\teff{$T_{\rm eff}$\,}
\newcommand{\angstrom}{\mbox{\normalfont\AA}}

\title[N spreads in $\sim$2 Gyr old LMC clusters]{On the Nitrogen variation in $\sim$2 Gyr old massive star clusters in the Large Magellanic Cloud\thanks{Based on observations collected at the European Organisation for Astronomical Research in the Southern Hemisphere under ESO programme(s) 0103.D-0248 (P.I. Martocchia), 088.D-0807(A) (archival data) and the VMC survey data.}}

\author[Martocchia S. et al.]{S. Martocchia$^{1}$\thanks{E-mail: s.martocchia@astro.ru.nl},
C. Lardo$^{2}$,
M. Rejkuba$^{3}$,
S. Kamann$^{4}$,
N. Bastian$^{4,5,6}$,
S. Larsen$^{1}$,
I. Cabrera-Ziri$^{4}$,
\newauthor W. Chantereau$^{7}$,
E. Dalessandro$^{8}$,
N. Kacharov$^{9}$,
M. Salaris$^{4}$.
\\
$^{1}$Department of Astrophysics/IMAPP, Radboud University, P.O. Box 9010, 6500 GL Nijmegen, The Netherlands\\
$^{2}$Dipartimento di Fisica e Astronomia, Universit\`a degli Studi di Bologna, Via Gobetti 93/2, 40129, Bologna, Italy\\
$^{3}$European Southern Observatory, Karl-Schwarzschild-Stra\ss e 2, D-85748 Garching bei M\"unchen, Germany\\
$^{4}$Astrophysics Research Institute, Liverpool John Moores University, 146 Brownlow Hill, Liverpool L3 5RF, UK\\
$^{5}$Donostia International Physics Center (DIPC), Paseo Manuel de Lardizabal, 4, 20018, Donostia-San Sebasti\'an, Guipuzkoa, Spain\\
$^{6}$IKERBASQUE, Basque Foundation for Science, 48013, Bilbao, Spain\\
$^{7}$Universit\'e de Strasbourg, CNRS, Observatoire astronomique de Strasbourg, UMR 7550, F-67000 Strasbourg, France\\
$^{8}$INAF-Osservatorio di Astrofisica \& Scienza dello Spazio, via Gobetti 93/3, I-40129, Bologna, Italy\\
$^{9}$Leibniz-Institut fuer Astrophysik, An der Sternwarte 16, 14482 Potsdam, Germany}

\date{Accepted XXX. Received YYY; in original form ZZZ}

\pubyear{2021}

\begin{document}
\label{firstpage}
\pagerange{\pageref{firstpage}--\pageref{lastpage}}
\maketitle

\begin{abstract}
We present ESO/VLT FORS2 low resolution spectroscopy of red giant branch stars in three massive, intermediate age ($\sim 1.7-2.3$ Gyr) star clusters in the Large Magellanic Cloud. We measure CH and CN index bands at 4300\AA\, and 3883\AA, as well as [C/Fe] and [N/Fe] abundance ratios for 24, 21 and 12 member stars of NGC~1978, NGC~1651, NGC~1783, respectively. We find
a significant intrinsic spread in CN in NGC~1978 and NGC~1651, a signal of multiple stellar populations (MPs) within the clusters. 
On the contrary, we report a null CN spread in NGC~1783 within our measurement precision.
For NGC~1978, we separated the two populations in the CN distribution and we translated the CN spread into an internal N variation $\Delta$[N/Fe]$=0.63\pm0.49$ dex. For NGC~1651 and NGC~1783,
we put upper limits on the N abundance variations of $\Delta$[N/Fe]$\leq 0.2, 0.4$ dex, respectively. 
The spectroscopic analysis confirms previous results from HST photometry, where NGC~1978 was found to host MPs in the form of N spreads, while slightly younger clusters (e.g. NGC~1783, $<$ 2 Gyr old) were not, within the limits of the uncertainties. It also confirms that intermediate age massive clusters show lower N abundance variations with respect to the ancient globular clusters, although this is in part due to the effect of the first dredge up at these stellar masses, as recently reported in the literature. We stress the importance of future studies to estimate the \textit{initial} N abundance variations, free of stellar evolutionary mixing processes, by observing unevolved stars in young clusters. 
\end{abstract}




\section{Introduction}\label{sec:intro}

One of the current outstanding problems in astrophysics is how globular clusters (GCs) form. A successful GC formation theory needs to reproduce the observed star-to-star chemical inhomogeneities in GCs that are often called ``chemical anomalies''.
Namely, all GCs that are massive enough ($\sim$ a few times $10^3$ \msun , \citealt{milone17})  
host multiple populations (MPs) of stars within them, characterised by one group of stars with the same chemical composition of the field (at similar metallicities) and another population (or more) having enhanced N, Na, He content, but depleted C and O (e.g. \citealt{gratton12}). The latter is typically called the anomalous or second population (2P), as opposed to normal or first population (1P). From both observational and theoretical sides, there is an ongoing
effort trying to unveil how MPs form. However, to date none of the scenarios put forward is fully satisfactory
(e.g. \citealt{bastianlardo18}).

Important characteristics of such chemical anomalies are: (i) they are found in all ancient and massive GCs (with
maybe one exception, Ruprecht 106, \citealt{dotter18,frelijj21}); (ii) they are present in all nearby galaxies where it was possible to probe them, such as the Magellanic Clouds, (MCs, e.g.  \citealt{mucciarelli09,dalessandro16,niederhofer17a,gilligan19}), M31 (e.g. \citealt{schiavon13,colucci14,sakari16}), the Fornax dwarf spheroidal (e.g. \citealt{larsen12,larsen14,martocchia20a}) and the Sagittarius dwarf (e.g. \citealt{carretta10,ftrincado21});
(iii) they are only found in high-density environments, i.e. they are marginally present in field stars (e.g. \citealt{martell11}) and have not been detected
in open clusters (e.g. \citealt{ocs}), (iv) the different populations form concurrently in age (within $\sim 20$ Myr, e.g. \citealt{nardiello15,martocchia18b,saracino20}),
(v) commonly the anomalous population is more centrally concentrated than the ``normal'', field-like population, although this seems to depend on the dynamical age of the clusters (e.g. \citealt{dalessandro19}). These are just a few of the peculiarities that a MP formation/evolution model has to satisfy. For more, we refer the interested readers to recent reviews such as \cite{bastianlardo18,gratton19}. 

Having determined their main characteristics, recently more studies have been dedicated to establishing
the behaviour of MPs as a function of cluster parameters with the aim to
provide fundamental constraints for any scenarios proposed for their origin. 
The mass of the cluster definitely plays a role in the onset of MPs, as it has been observed that higher mass clusters show higher fractions of 2P stars, along with higher He, N and Na variations \citep{schiavon13,carretta14,milone17,milone18,lagioia19}. 
Furthermore, the search for MPs has been expanded in star clusters that have similar masses, but are much younger than the ancient GCs, down to $\sim$600 Myr \citep{bastian20}. 
\cite{niederhofer17a, niederhofer17b,martocchia18a,martocchia19} found that chemical anomalies, in the form of N spreads, are present in massive star clusters of the Magellanic Clouds (MCs) that are older than $\sim$ 2 Gyr, while none are found in clusters younger than this age \citep{martocchia17,zhang18}. 
Finding MPs in clusters as young as 
$\sim$2 Gyr 
implies that the chemical variations must form through mechanisms that acted until the present day and are not due to
the special conditions of the early Universe. 

The presence of MPs in intermediate age clusters has been also established through spectroscopic studies at low-resolution, aimed at revealing intrinsic N variations within the clusters \citep{hollyhead17,hollyhead18, hollyhead19}.
Chemical anomalies in intermediate age massive star clusters were only found in the form of N spreads, and associating these phenomena with 
those observed in the ancient GCs was still quite tentative. However, recent studies bolstered the idea that young and ancient star clusters are just the same type of objects seen at different stages of their lifetimes. 
It has been lately demonstrated
that young star clusters show variations in elements other than N, following the same chemical patterns observed in old GCs, e.g.\, 
signatures of He variations within intermediate-age MCs star clusters  \citep{chantereau19,lagioia19}. Na variations have also been found, by first using integrated light techniques in quite young clusters ($\sim$2-3 Gyr, \citealt{bastian19}).
Subsequently, \cite{saracino20b} and \cite{martocchia20b} showed that Na variations are present in massive star clusters at different ages ($\sim 2-7.5$ Gyr), by combining the power of HST and VLT/MUSE, a
technique introduced by \cite{latour19}.  

\begin{figure}
\centering
\includegraphics[scale=0.64]{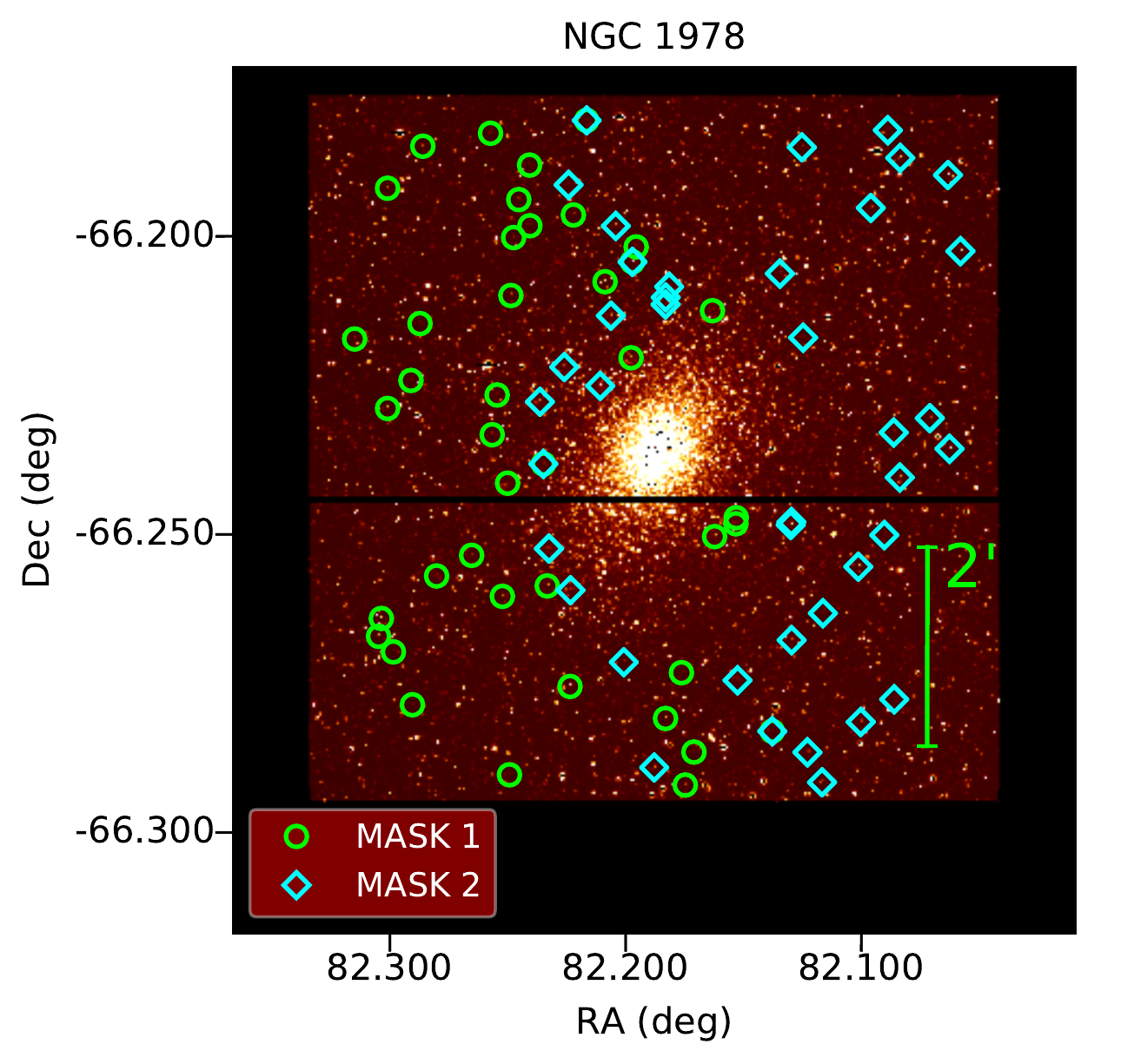}
\caption{FORS2 $v_{HIGH}$ mosaic image of NGC~1978 field of view. The green circles and cyan diamonds indicate targeted stars for the MASK 1 and MASK 2 of this cluster, respectively. The green bar on the lower right side denotes the projected distance of 2 arcmin.}
\label{fig:fov}
\end{figure}

Additionally, the magnitude of the N spread (from photometric colour spreads) detected in individual clusters is found to be positively correlated with the age of the clusters \citep{martocchia19}. Part of such a correlation is most likely due to the effect of the first dredge-up, FDU, at different ages \citep{salaris20}.
However, the FDU cannot fully explain the observed correlation. This dependence on age is not predicted by any model that has been proposed to explain the formation and evolution of MPs and its origin is still under investigation. 
It is important to confirm this result using spectroscopy to directly assess the level of N variations (if any) within young/intermediate age stellar systems.

In this paper, we report a spectroscopic ESO VLT/FORS2 study of red giant branch (RGB) stars of three intermediate age massive star clusters in the Large Magellanic Cloud (LMC), $\sim 1.7-2.3$~Gyr old, namely NGC~1651, NGC~1783, NGC~1978. High resolution spectroscopic studies of these three clusters did not find evidence for Na and O variations within them \citep{mucciarelli08}, while recently \cite{saracino20b} showed the presence of small Na variations in NGC~1978.

Here we report the measurements of CH and CN indices, as well as [C/Fe] and [N/Fe] abundance ratios.
Also, NGC~1978 was photometrically found to host MPs in the form of N spreads in our HST survey, while NGC~1783 did not \citep{martocchia18a}.
No studies of MPs have instead been carried out for NGC~1651 so far, to the best of our knowledge.

The paper is structured as follows: Section \ref{sec:obs} describes the observations and data reduction procedures. Section \ref{sec:analysis} reports on the spectral analysis. 
In Section \ref{sec:res} we outline the results, while we discuss and conclude in Section \ref{sec:disc}.

\begin{table*}
\centering
\caption{FORS2 observation log and main properties of the clusters presented in this paper. Columns report the following information: name of the cluster, mask number, exposure times, total number of stars targeted, mean SNR with 1$\sigma$ dispersion, age of the cluster, mass of the cluster, metallicity of the cluster [Fe/H], distance modulus $(m-M)$, extinction coefficient $A_V$, stellar mass of a typical RGB star of the cluster $M_{\star}$, half light radius $r_h$. 
}
\label{tab:infolog}
\begin{tabular}{l c c c c c c c c c c c}
\hline
Cluster & Mask & Exposures & $N_{\rm stars}$ & $<SNR>\pm\sigma$ & Age & Mass  & [Fe/H] & $(m-M)$ & $A_V$ & $M_{\star}$ & $r_h$ \\
        & No. & (s)& & & (Gyr) & ($\times10^5$\msun) & (dex) & (mag) & (mag) & (\msun) & ($\arcmin$)\\
\hline
NGC~1783 & 1      & 4$\times$3300, 2$\times$2000 & 39 & $19.4\pm7.8$ & 1.7 & 2.5$^1$  & $-0.40$ & 18.49 & 0.00 & 1.59 & 0.78$^1$\\
\hline
NGC~1651 & 1      & 4$\times$3300, 1$\times$1500 & 32 & $40.7\pm10.3$ & 2.0 & 0.8$^1$ & $-0.30$  & 18.42 & 0.15 & 1.55 & 0.88$^1$ \\
         & 2      & 4$\times$3300 & 37 & $32.8\pm10.6$ & & & & & & &\\
\hline
NGC~1978 & 1      & 3$\times$3300, 1$\times$3600 & 41 & $28.2\pm11.7$ & 2.3 & 2-4$^2$ & $-0.35$  & 18.55 & 0.16 & 1.50 & 0.52$^3$ \\
         & 2      & 5$\times$3300 & 40 & $27.0\pm9.0$ & & & & & & & \\
\hline
\end{tabular}
\\
$^1$~\citet{goudfrooij14},
$^2$~\citet{westerlund97},
$^3$~\citet{dalessandro19}.
\end{table*}

\section{Observations and Data Reduction}\label{sec:obs}

Data for NGC~1783, NGC~1651 and NGC~1978 were obtained with FORS2 \citep{appenzeller}, mounted on the Cassegrain focus of UT1 of the ESO VLT/Paranal observatory in Chile (programme ID 0103.D-0248, P.I. Martocchia). The  instrument was used in the so-called mask exchange unit (MXU) mode with laser-cut invar masks inserted in the focal plane allowing acquisition of spectra for $\sim 40$ targets spread over $6.8' \times 6.8'$ field of view. 
We observed two masks for NGC~1978 and NGC~1651 and one mask for NGC~1783. For each mask, the upper CCD (chip1) was centred approximately on the centre of the cluster, while the lower CCD (chip2) pointed southwards from the cluster centre.
We used the 600B+22 grism to sample the spectral region where the CN ($\sim$3883\AA) and CH ($\sim$4300\AA) features are located.
The typical resolution of the spectra is $R=\lambda/\Delta \lambda \simeq 800$ (around $\lambda=4627$\AA) while the nominal spectral coverage is  $\sim3300 - 6210$\AA. However, depending on the location of the slit in the mask, some stars had a different spectral coverage\footnote{We note that this is not affecting the analysis or results of the paper, because we took care that all targets have spectra starting from $\sim$3500\AA\, and thus including the blue CN band. On the red side some targets have spectra extending only up to $\sim$5200\AA.}.

The observations were taken in Visitor Mode over 5 half-nights from September 25th to 29th, 2019. After the acquisition, the mask was centred by taking through-slit images in which our targets were suitably exposed. Due to observations at relatively high airmass, we checked after typically every 2 exposures that the mask was still well centred by taking another through-slit image.

Around 30-40 slits per mask (with width of 1 \arcsec\, and variable length of 6 to 10\arcsec) were allocated for each cluster. In this way, we obtained data for more than 80 stars for NGC~1978, $\sim$70 for NGC~1651 and $\sim$40 stars for NGC~1783. 
Exact numbers are reported in Table \ref{tab:infolog} along with the exposures and information about each cluster.

Primary targets were selected from the RGB area of each cluster using FORS2 photometry.
Where it was impossible to position a slit on a primary target, a random star was chosen in its place.
Also, in a few cases, the slit length of the RGB stars was increased to allow more background to be sampled. The images for NGC~1978 and NGC~1783 were acquired as part of the programme presented in this paper, while we used available FORS2 images from the ESO archive (programme ID 088.D-0807A) for NGC~1651. We report on how the pre-imaging was reduced in the next Section \ref{subsec:phot}.
Figure \ref{fig:fov} shows the FORS2 mosaic image used to select targets in NGC~1978, as an example. Green open circles and cyan open diamonds indicate selected target stars for MASK 1 and MASK 2, respectively. It is possible to appreciate the size of the cluster core with respect to the FORS2 field of view. Due to the small spatial extent and very high crowding, it was not possible to select targets close to the centres of the clusters. Besides using color-magnitude diagrams (CMDs) to select spectroscopic targets, the target selection also included visual inspection of pre-imaging data and excluded stars with (similar brightness or brighter) neighbours within at least 2\arcsec\, of the slit location.

We will describe the spectroscopic reduction in Section \ref{subsec:spectro}.

\begin{figure*}
\centering
\includegraphics[scale=0.55]{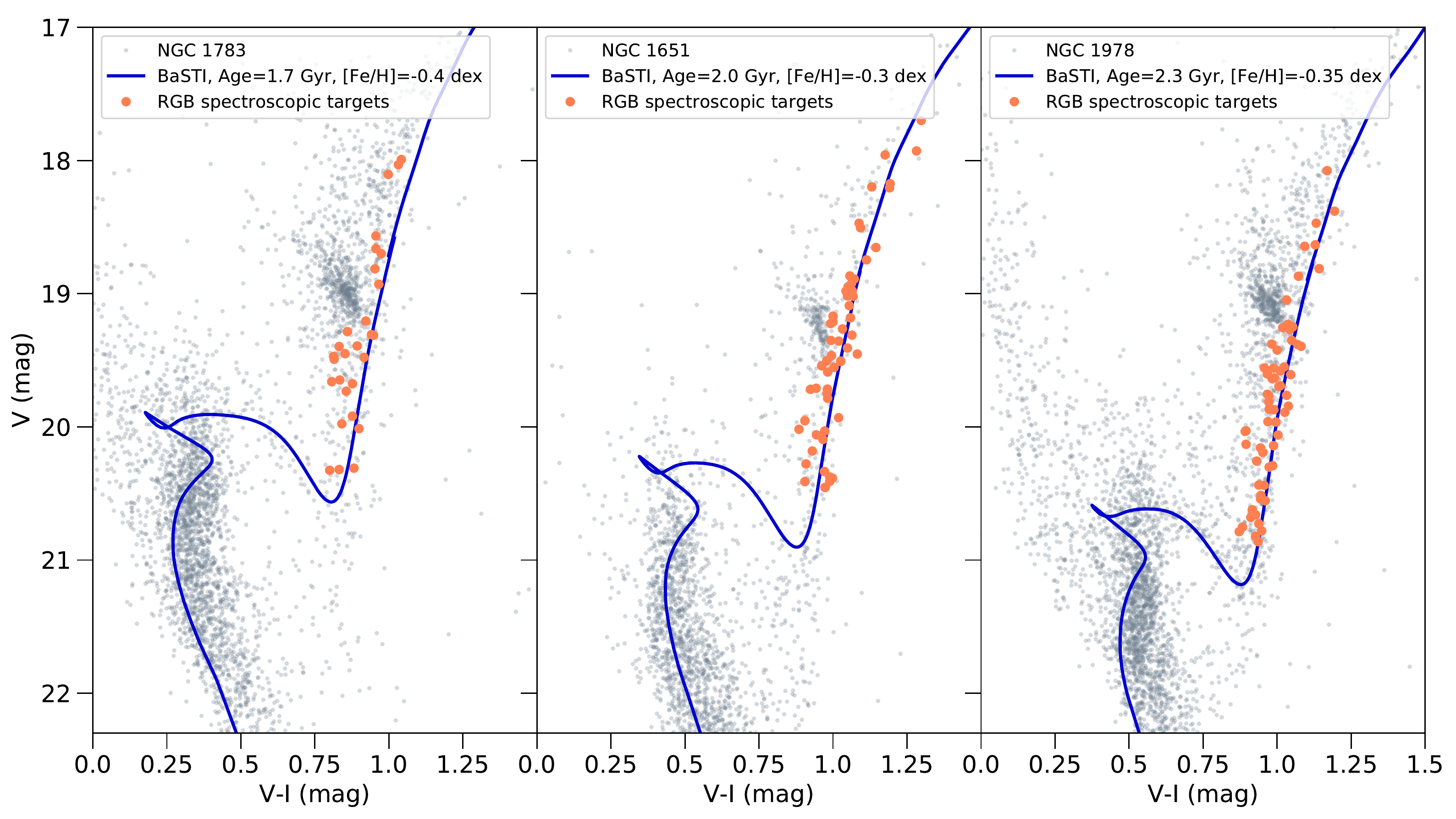}
\caption{CMDs of the FORS2 pre-imaging (for stars with distance $<2.5$ arcmin from the cluster centre) in $V-I$ vs $V$ Johnson-Cousins filters for the three clusters presented in this work. Orange circles indicate RGB spectroscopic targets in each panel. The blue curve indicates the BaSTI isochrone in each panel for values of age and metallicity reported in the legend. For the values of distance modulus and extinction, see Table \ref{tab:infolog}.}
\label{fig:cmds}
\end{figure*}
 
\subsection{Pre-imaging and photometry}\label{subsec:phot}

We obtained FORS2 pre-imaging data for NGC~1978 and NGC~1783 using $v_{HIGH}$ and $I_{BESS}$ filter observations centred on the same pointings as for the spectroscopic follow-up observations with FORS2 MXU.
The pre-imaging included 3$\times$30s short exposures for each filter and 3$\times$300s and 3$\times$180s long exposures for the $v_{HIGH}$ and $I_{BESS}$ filters, respectively. 
For NGC~1651, $v_{HIGH}$, $I_{BESS}\,$ images were instead already available in the archive. They had 8$\times$30s exposures each. 

The images were processed, flat-field corrected, and bias-subtracted
using the FORS2 pipeline in the ESO Reflex data processing environment \citep{ESOreflex}. 

The photometric catalogues have been obtained following a similar strategy as in \cite{martocchia19}, from \cite{dalessandro14, dalessandro18}. 
We used DAOPHOTIV \citep{stetson87} independently on each filter and each chip. We selected several hundreds of bright and isolated stars in order to model the point-spread function (PSF). All available analytic functions were considered for the PSF fitting (Gauss, Moffat, Lorentz and Penny functions), leaving the PSF free to spatially vary to the first-order. In each image, we then fit all the star-like sources detected at $3\sigma$ from the local background with the best-fit PSF model by using ALLSTAR. We then created a master catalogue composed of stars detected in (n/2 +1) images for each cluster\footnote{Where the number of exposures in the same filter is equal to three, we used stars detected in 2 images to create the catalogues.}. The final star lists for each image and chip were cross-correlated by using DAOMATCH, then the magnitude mean and standard deviation measurements were obtained through DAOMASTER. We obtained the final catalog by matching the star lists for each filter by using DAOMATCH and DAOMASTER.

We matched our photometric catalogues with the \cite{zaritsky04} catalogues, to convert instrumental magnitudes to the Johnson-Cousins photometric system $V$ and $I$, and instrumental coordinates to the absolute image World Coordinate System. This was performed by using CataXcorr\footnote{Part of a package of astronomical softwares (CataPack) developed by P. Montegriffo at INAF-OABo.}.
Figure \ref{fig:cmds} shows the CMDs of the three clusters in $V-I$ vs. $V$ Johnson-Cousin filters for stars that are within 2.5 arcmin from the cluster centre. This number was just chosen to minimize the inclusion of field stars in the plot. It roughly corresponds to 3, 3 and 5 times the half light radius ($r_h$, see Table \ref{tab:infolog}) for NGC~1651, NGC~1783 and NGC~1978, respectively. 
The analysed RGB spectroscopic targets are superimposed as orange circles. 

We complemented our optical photometry with near-infrared (NIR) ESO VISTA Magellanic Cloud (VMC) photometric survey data \citep{cioni11} based on observations in 
$Y$, $J$ and $Ks$ VIRCAM filters. Details about the VMC PSF photometry 
can be found in \cite{rubele15}. $J$ and $Ks$ VISTA magnitudes were transformed to the 2MASS photometric system by using the relations reported in \cite{vista2mass}. The VMC photometry was used
for membership determination using the position of the stars on different CMDs based on independent datasets (see Sect. \ref{subsec:mem}) and for atmospheric parameters estimation (see Sect. \ref{subsec:params}.)

\subsection{Spectroscopy}\label{subsec:spectro}

Spectra were reduced running the FORS2 pipeline in the ESO Reflex data processing environment. This included bias frames subtraction, flat field normalisation, wavelength calibration and the 1D spectra extraction for each exposure.
Single exposures were median combined with the \textit{scombine}
\texttt{IRAF} routine for each star. 
Hot stars characterised by prominent Balmer lines in their spectra 
were rejected from the following analysis. These are main-sequence (MS) stars that were randomly obtained from the masks, where it was impossible to position a slit on a primary target (see Sect. \ref{sec:obs}). This rejects one star for NGC~1978 and one for NGC~1651.

We applied Doppler correction to bring the spectra to the reference/laboratory reference wavelength frame. We report on how the RVs were estimated in Appendix \ref{sec:rv}. 
Final values for the RVs of individual stars\footnote{We note that these are the actual measured RVs of the stars and not the RVs used to shift the spectra to zero velocities, see Appendix \ref{sec:rv} for more details.} are reported in Table \ref{tab:info_spectra}.

Given the low resolution of the spectra and consequently large uncertainties on the RVs, as well as given the systematic offsets described in Appendix \ref{sec:rv}, we were unable to perform a membership analysis based on the kinematics of the stars. Other criteria were used, as reported in the next Section \ref{subsec:mem}.

For the following analysis, we considered stars with SNR $>5$ in the CN(3883\AA) region. This data quality selection removes 1 star for NGC~1651, 2 stars for NGC~1978 and 10 stars for NGC~1783. 
\begin{figure*}
\centering
\includegraphics[scale=0.64]{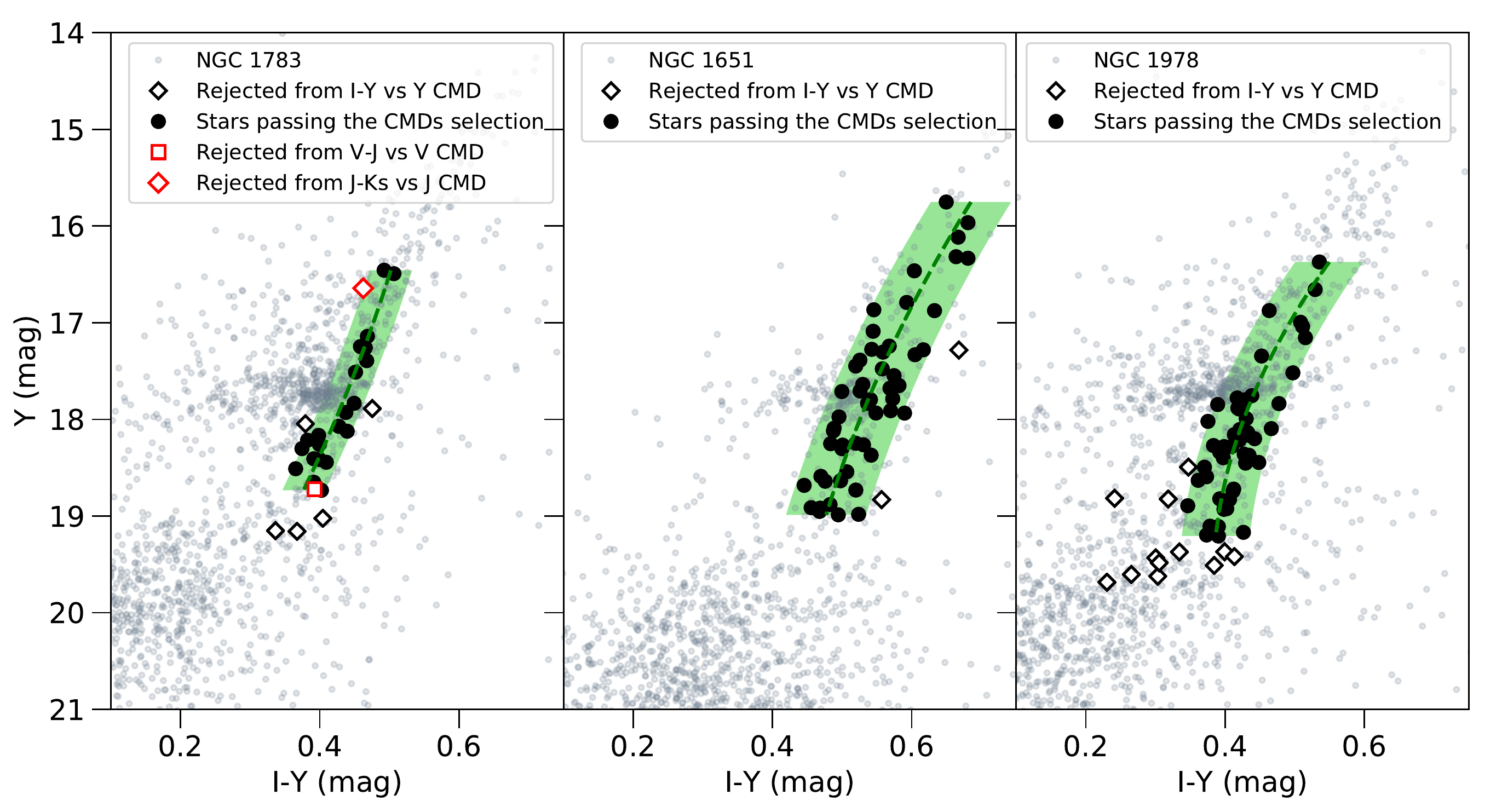}
\caption{$I-Y$ versus $Y$ CMDs for NGC 1783 (left panel), NGC 1651 (middle panel) and NGC 1978 (right panel). Black filled circles indicate stars that pass the CMDs selection. White diamonds with black contours represent stars that did not pass the $I-Y$ vs $Y$ CMD selection. for NGC~1783, the red and white square indicates a star that did not pass the $V-J$ vs $V$ CMD selection, while the red and white diamond is a star that did not pass the $J-Ks$ vs $J$ CMD selection.  The green dashed lines show a fiducial line to the data with $Y<19$ mag, while the green shaded area represents the 2$\sigma$ dispersion around the fiducial lines. See text for more details.} 
\label{fig:mem_cmd_iy}
\end{figure*}
\begin{figure*}
\centering
\includegraphics[scale=0.41]{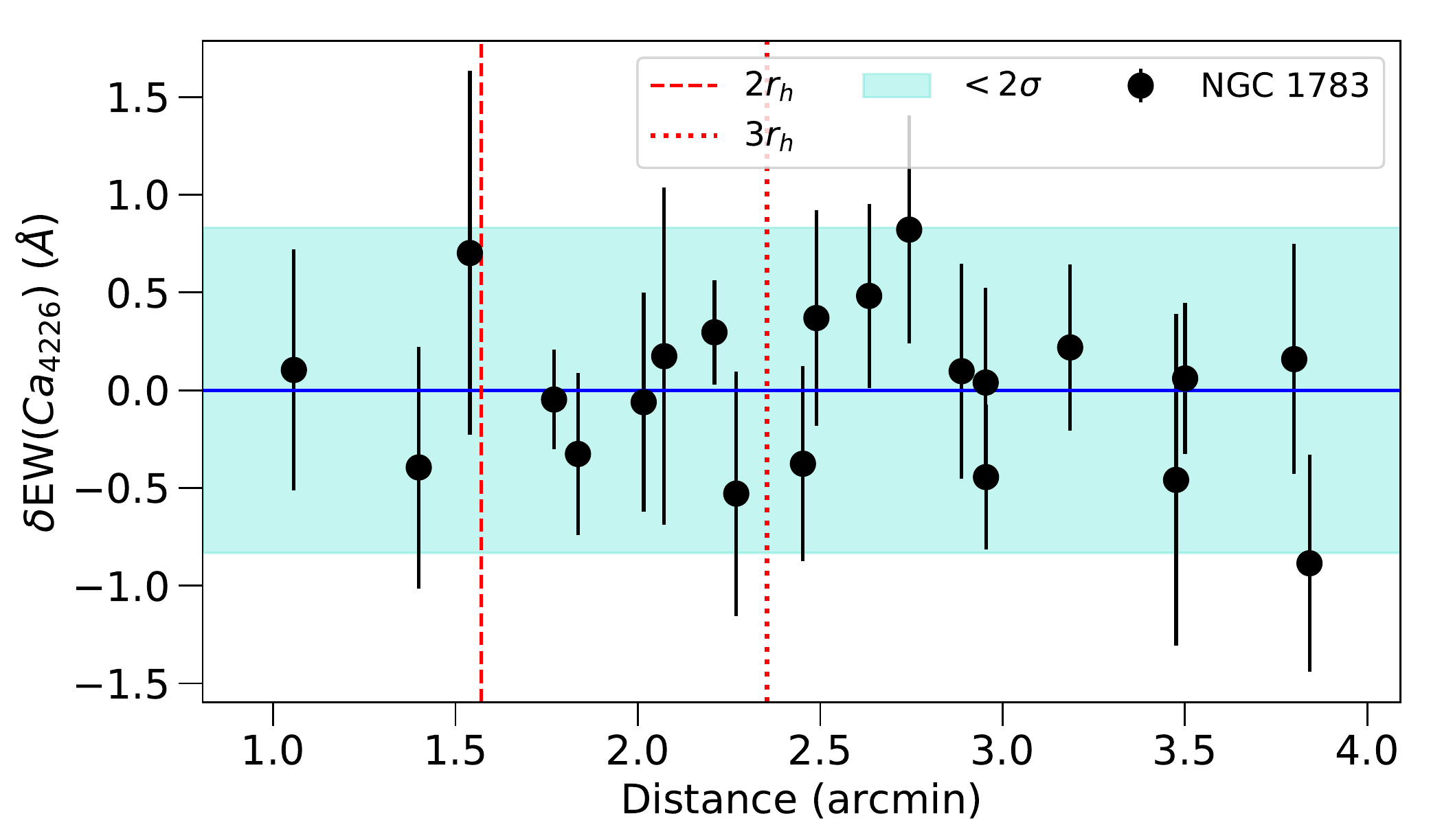}
\includegraphics[scale=0.41]{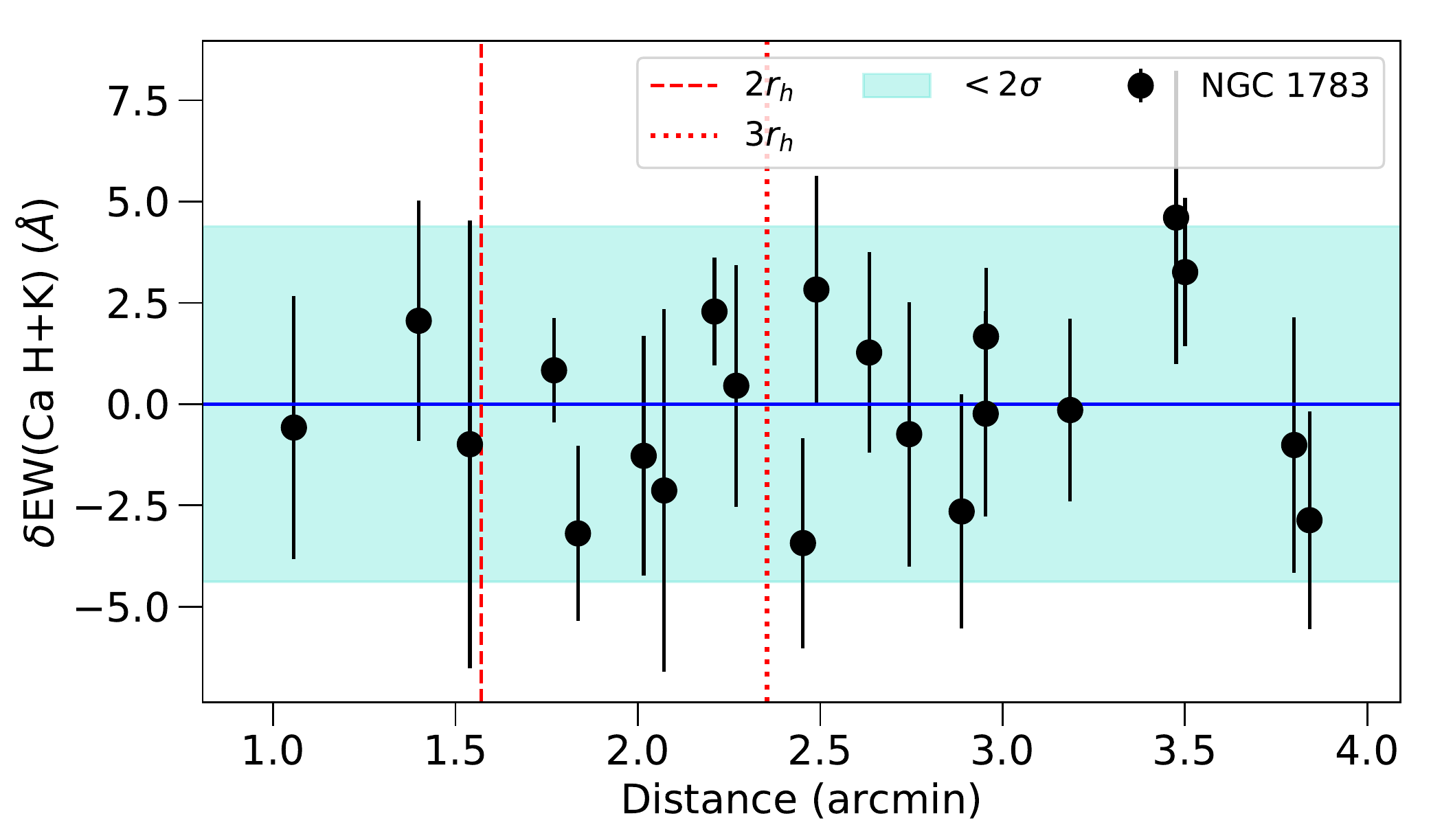}
\includegraphics[scale=0.41]{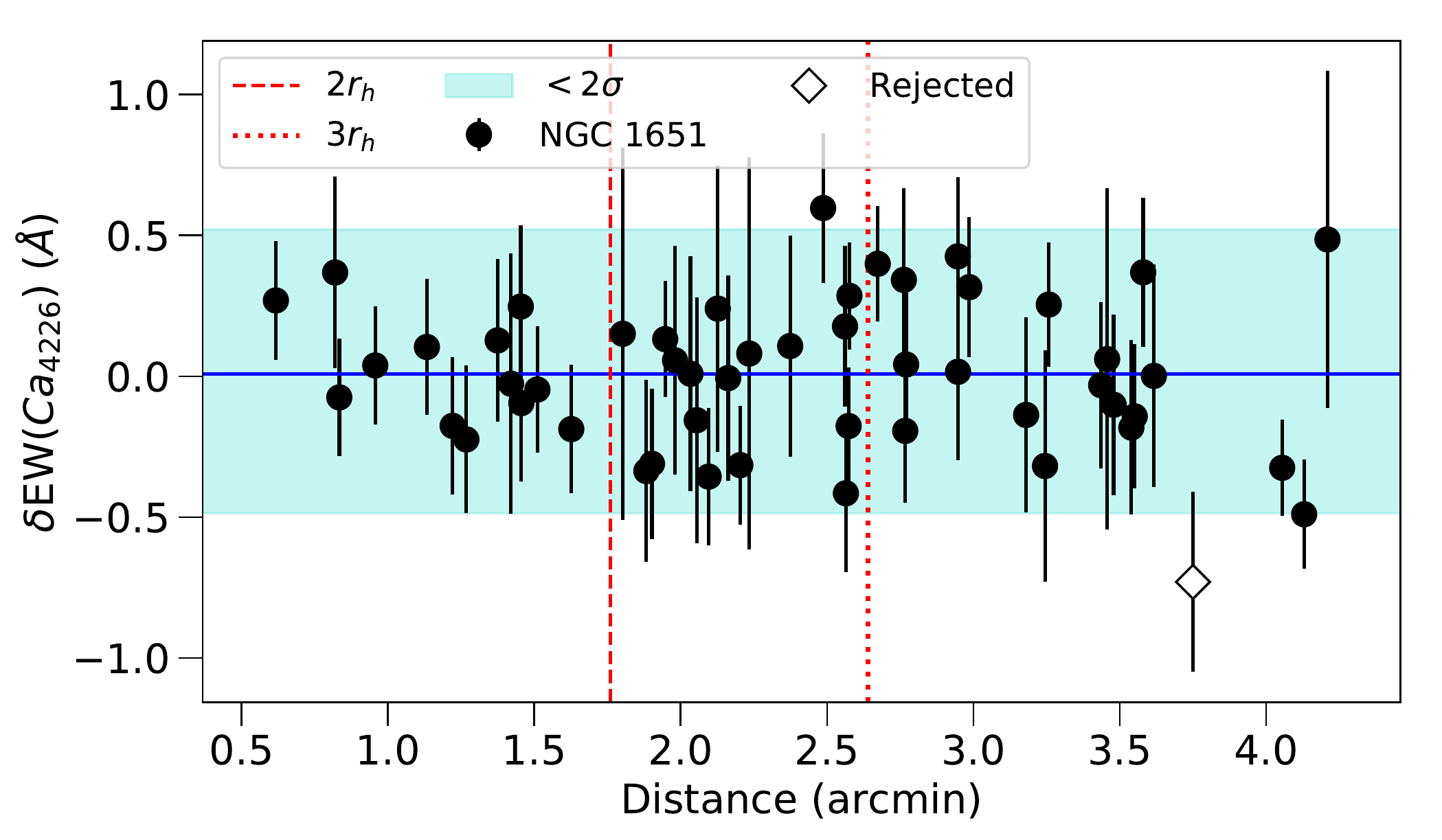}
\includegraphics[scale=0.41]{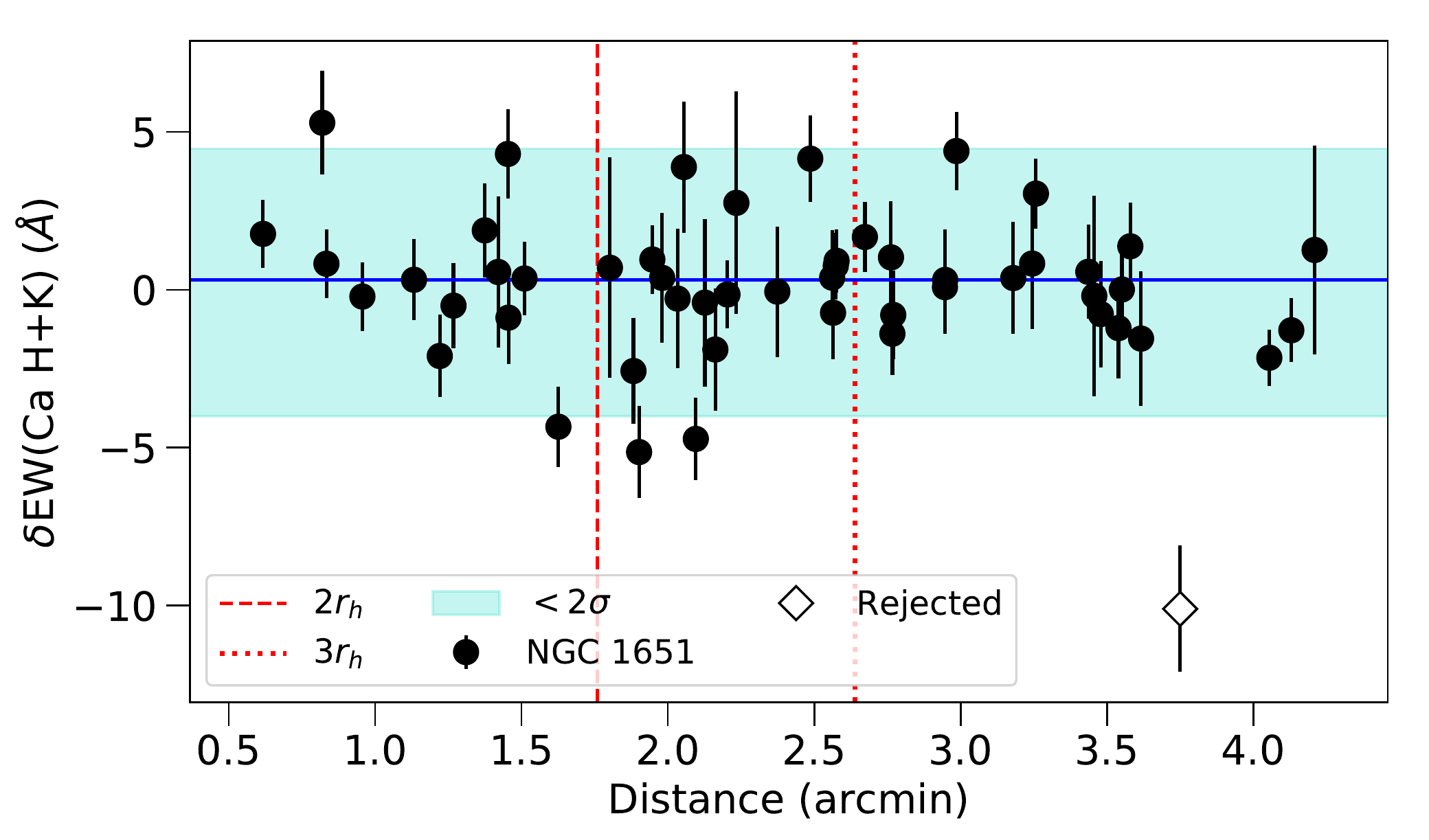}
\includegraphics[scale=0.41]{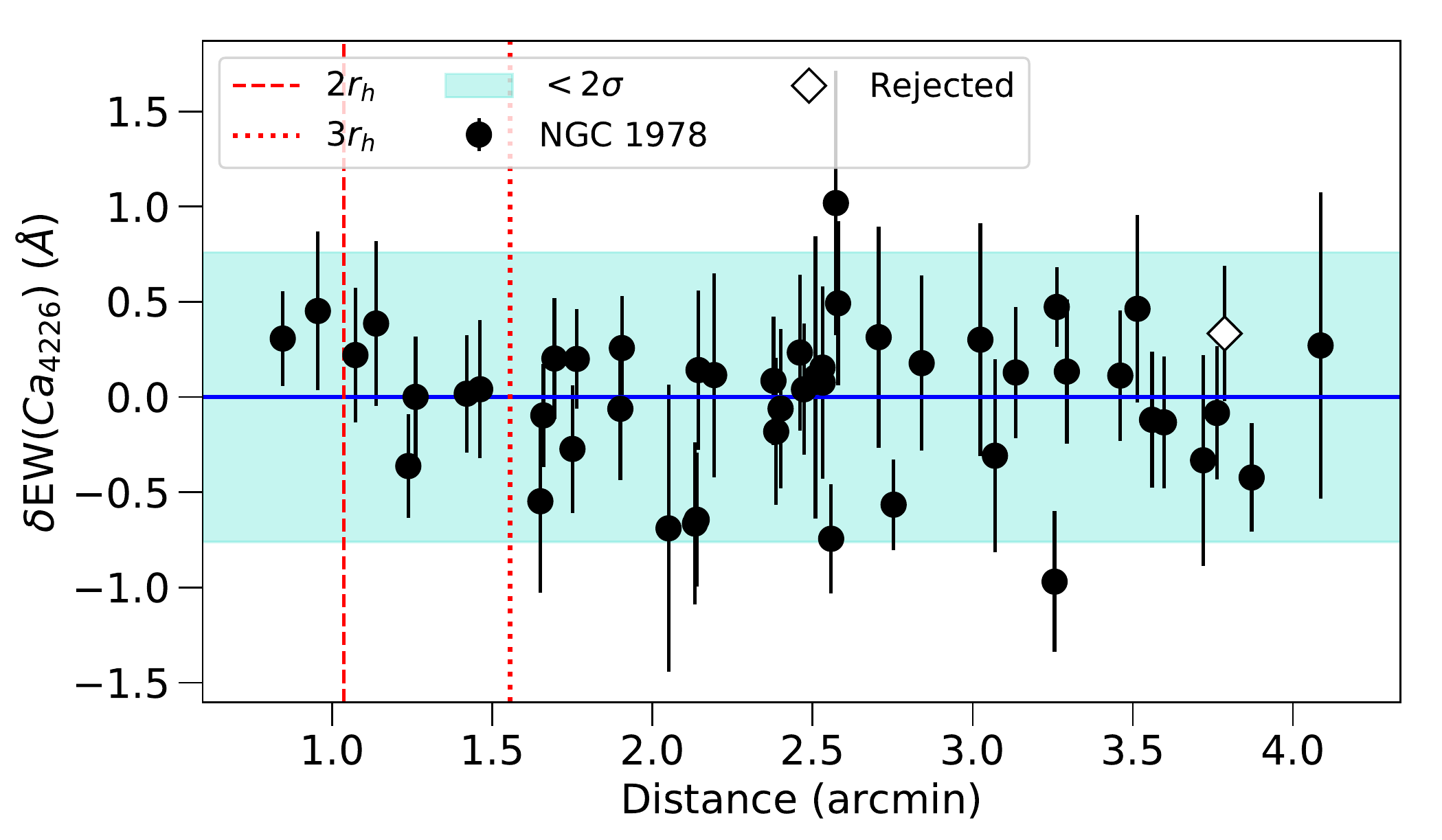}
\includegraphics[scale=0.41]{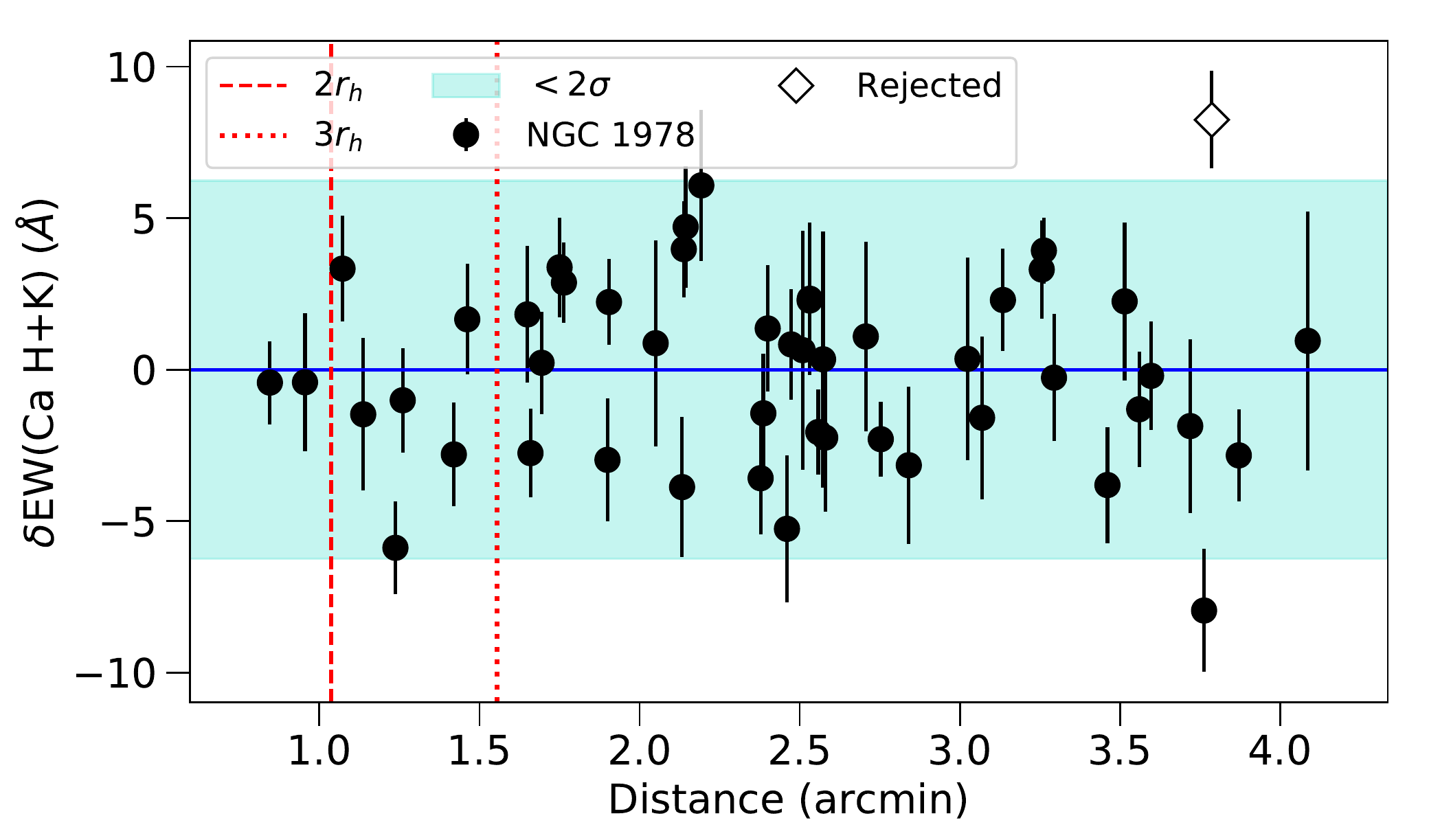}
\caption{EW of Ca at 4226\AA (left panels) and Ca (H+K) (right panels) lines after they have been corrected for luminosity dependence as a function of the distance from the cluster centre, for each cluster in our sample. Black circles represent stars that pass the selection, while white diamonds represent rejected stars. Blue horizontal lines show mean values, whereas cyan shaded areas denote the 2$\sigma$ dispersion around the mean. Vertical red dashed and dotted lines indicate 2 and 3 times the half light radius, respectively. See text for more details.}
\label{fig:mem_ew_ca}
\end{figure*}

\begin{figure*}
\centering
\includegraphics[scale=0.51]{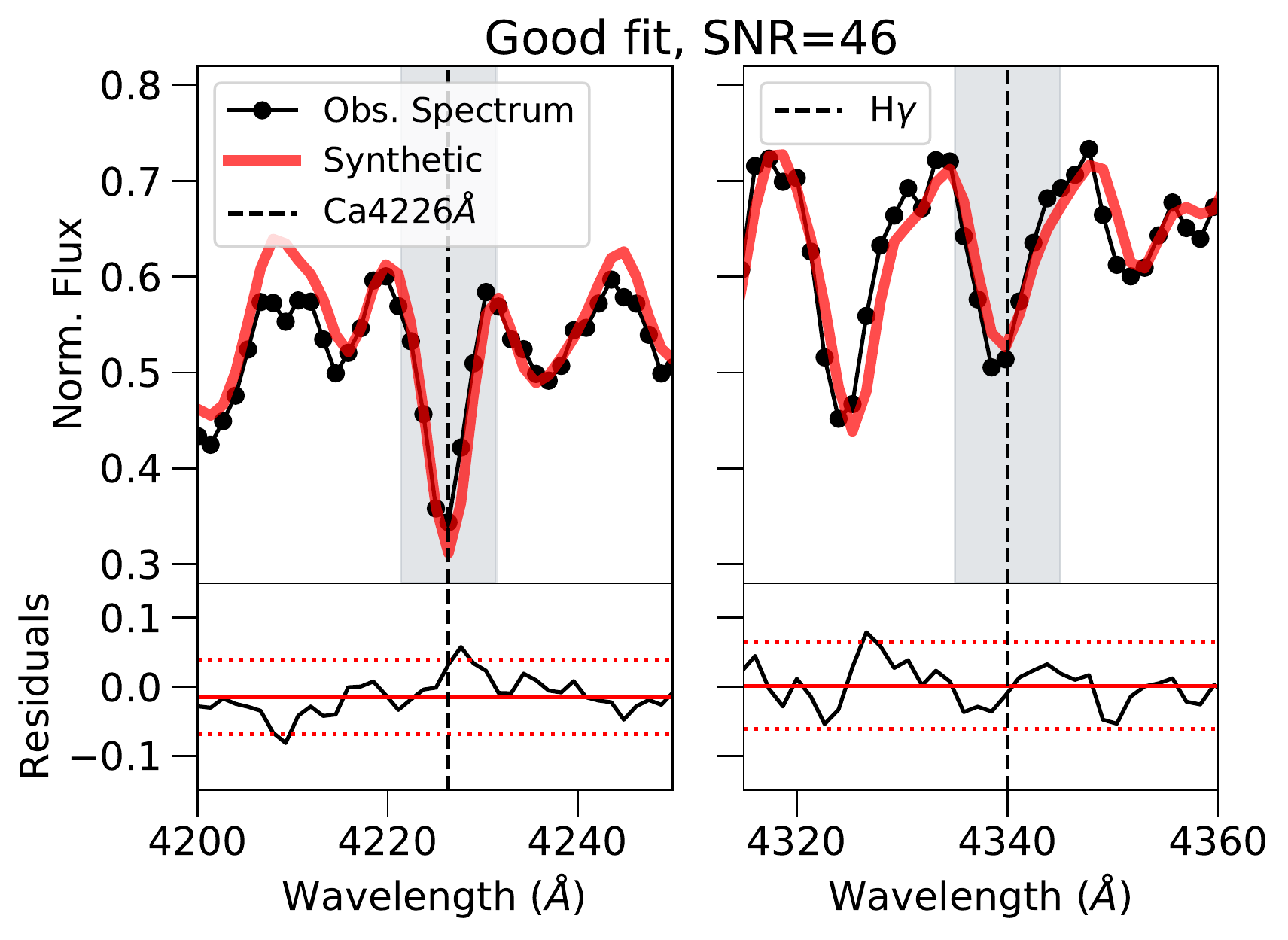}
\includegraphics[scale=0.51]{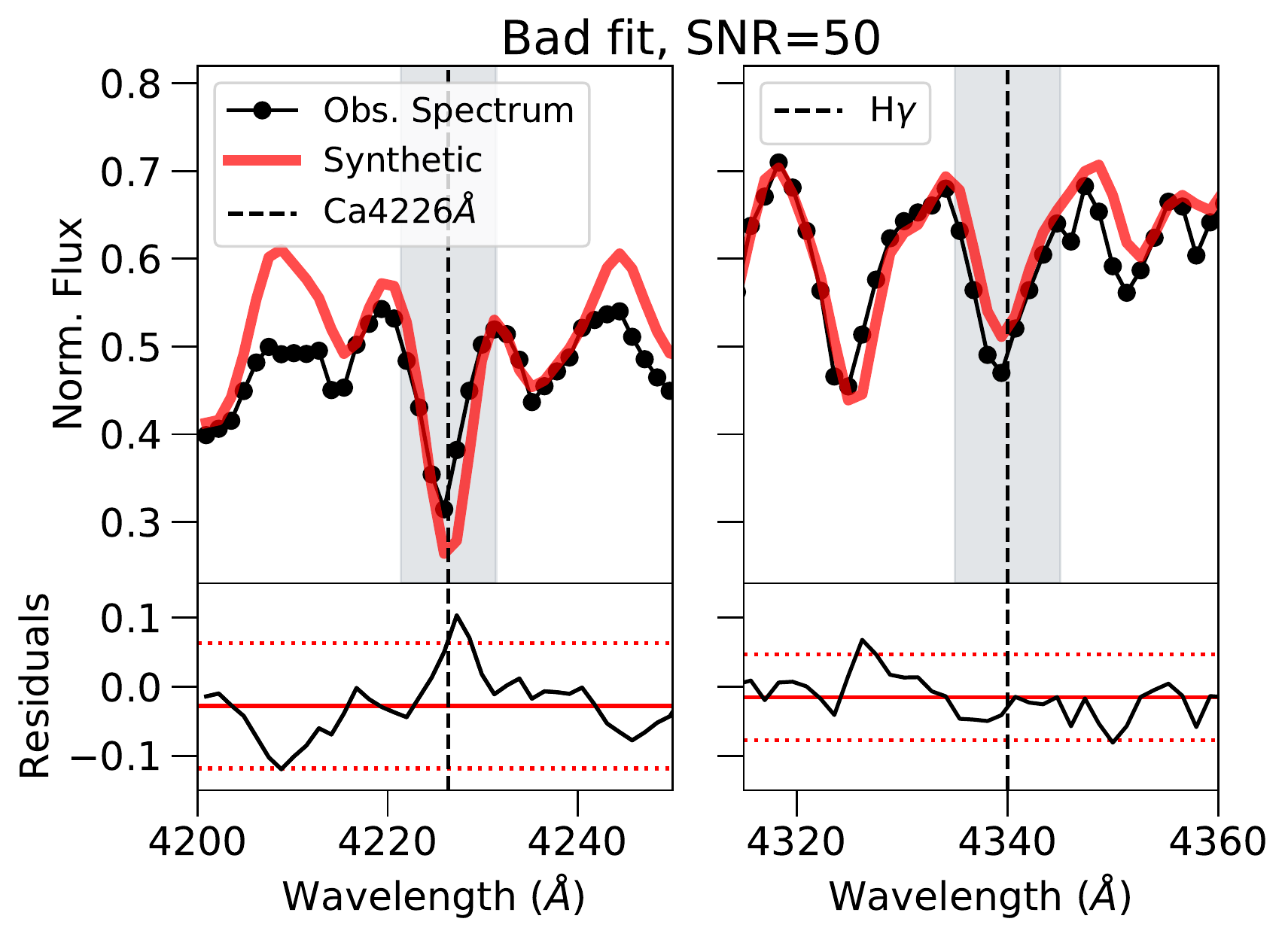}
\caption{Left panel: observed (black) and synthetic (red) spectrum for a member star of NGC~1651 (N1651-34), showing the example of a good fit around the Ca(4226\AA) and H$\gamma$ lines. The line centres are indicated using dashed vertical lines. The grey shaded area represents a region of a 10\AA\, width where the comparison between the observed and synthetic spectra was performed. Residuals of the measured difference between the observed spectrum and the theoretical model are plotted in the lower panel. Red horizontal solid and dotted lines indicate the mean and 2$\sigma$ dispersion on the residuals, respectively. 
Right panel: same as in the left panel but showing an example of a bad fit for a non member star of NGC~1651 (N1651-50), with similar SNR.}
\label{fig:mem_ca_fit}
\end{figure*}

\section{Spectral Analysis}\label{sec:analysis}
\subsection{Cluster Membership}\label{subsec:mem}

Besides the cluster stars, in the same region of the CMD there could be also red giants from the surrounding LMC field population, having similar colours due to an age-metallicity degeneracy. Fig. \ref{fig:fov} shows the FORS2 $v_{HIGH}$ mosaic image of NGC~1978 field of view (fov) considered for the selection of the targets and thus for the masks preparation. This is a 6.8$\arcmin \times 6.8\arcmin$ fov, centred on the centre of the cluster.  The fov can reach distances more than 5 times the half light radius of the clusters (see Table \ref{tab:infolog}). It is then important to select bona-fide cluster members. Member stars were selected according to both photometric and spectroscopic criteria.

Photometrically, we checked the positions of our targets in several different CMDs, coupling optical and NIR filters, to reject those stars that were not lying in the cluster RGB region. For each cluster, we first plotted our targets in the $I-Y$ vs. $Y$ CMD, then in the $V-J$ vs. $V$ CMD, in the $V-Ks$ vs $V$, the $J-Ks$ vs $J$ and finally in the $V-I$ vs $V$ CMD. Figure \ref{fig:mem_cmd_iy} shows the $I-Y$ vs. $Y$ CMDs, which were very useful to discriminate those stars lying very close (or on top) the subgiant branch (SGB) phase. Indeed, such probable SGB stars, at $Y\gtrsim 19$ mag, were rejected from the following analysis. For stars with $Y\lesssim 19$ mag, the green dashed lines in Fig. \ref{fig:mem_cmd_iy} show a fiducial line fit to the data, while the green shaded area represents the 2$\sigma$ dispersion around the fiducial line. Stars lying outside the 2$\sigma$ dispersion around the line were rejected. This was checked in every CMD reported above. In Fig. \ref{fig:mem_cmd_iy}, black filled circles indicate stars that pass the CMDs selection. White diamonds with black contours represent stars that did not pass the $I-Y$ vs $Y$ CMD selection. For NGC~1783, the red and white square indicates a star that did not pass the $V-J$ vs $V$ CMD selection, while the red and white diamond is a star that did not pass the $J-Ks$ vs $J$ CMD selection. In total, the photometric criterion removed 7 stars for NGC~1783, 2 stars for NGC~1651 and 12 stars for NGC~1978.

Additionally, we applied the following spectroscopic criteria to select cluster members:

\begin{enumerate}
    \item stars with discrepant equivalent width (EW) value for the Ca line at 4226\AA\, and Ca H+K lines were rejected;
    \item stars with a bad fit with synthetic spectra around the Ca(4226\AA) and H$\gamma$ lines were rejected.
\end{enumerate}

For the first criterion (i), we measured EWs for three Ca lines that are clearly visible in the spectra and that can be considered as a proxy for metallicity, i.e. the Ca line at 4226\AA\, and the Ca H+K lines. EWs were measured by using the \textit{splot} task in \texttt{IRAF}.
We removed the dependence of the EWs on the luminosity by doing a linear fit on Ca(4226\AA) and Ca(H+K) as a function of the $V$ magnitude. Then, we calculated the EW residuals with respect to the linear fit, thus obtaining a $\delta$EW(Ca4226\AA) and a $\delta$EW(Ca H+K).  
Figure \ref{fig:mem_ew_ca} shows the luminosity-corrected $\delta$EW(Ca4226\AA) and $\delta$EW(Ca H+K) as a function of the distance from the cluster centre. The blue horizontal line in each panel shows the mean value of the plotted quantities and the cyan shaded area indicates the 2$\sigma$ dispersion around the mean. The red dashed and dotted vertical lines represent the distance at 2 and 3 times the half light radius respectively, for each cluster. Unfortunately, in NGC~1978 we are sampling mainly the outskirts of the cluster, as the majority of our stars is $>3r_h$ distant from the centre. However, the tidal radius of our clusters is $>5\arcmin$ (e.g. \citealt{goudfrooij11,goudfrooij14}). 

Black filled circles in Fig. \ref{fig:mem_ew_ca} represent stars that pass this selection. According to this criterion, we kept all the stars that are consistent with the 2$\sigma$ dispersion, within the errors. As it is clear from the plots, the errors are quite large and only 2 stars are removed, one for NGC~1978 and one for NGC~1651\footnote{We additionally checked other lines such as Fe at 5015\AA, Fe at 5270\AA\, and Mgb but there is not a clear separation between cluster and field stars, as most likely a combination of low resolution and low SNR hampers the possibility to distinguish them effectively.}.
We note that these two stars are also rejected through the next criterion (ii).

For the criterion (ii), we calculated a synthetic template with the same parameters as the observed spectrum, around the region of Ca(4226\AA) and H$\gamma$ ($\sim$4340\AA) lines and we matched it with the observed spectrum. Details about how the stellar atmospheric parameters were estimated are reported in the next Section, \ref{subsec:params}, while we explain how synthetic spectra are calculated in Section \ref{subsec:ind}.
We observed that some fits were not good, meaning that
the assumed metallicity or the effective temperature of the synthetic spectra did not match the observation, i.e. very likely such stars are not members of the cluster\footnote{We did not attempt to measure the metallicity from the individual spectra because of the low resolution of the FORS2 data, at which Fe lines are blended. However, we note that the metallicity of member stars for all three clusters has been already measured through high resolution spectroscopy \citep{mucciarelli08}. We adopted a single metallicity value for each cluster as reported in Table \ref{tab:infolog}.}.

Figure \ref{fig:mem_ca_fit} shows the comparison between a good (left panels) and a bad fit (right panels) for two stars of NGC~1651, with a similar SNR. We report the observed (black) against the synthetic (red) spectrum around the Ca(4226\AA) and the H$\gamma$ lines, which are highlighted with dashed black lines. The grey shaded area indicates a region of a 10\AA\, width where the comparison between the observed and synthetic spectra was performed. Residuals of observed minus synthetic spectrum are reported in the lower panels. Red horizontal solid and dotted lines indicate the mean and 2$\sigma$ dispersion on the residuals, respectively. It is possible to note how the dispersion on the residuals of the bad fit is much larger than the one of the good fit. 

\begin{figure}
    \centering
    \includegraphics[scale=0.51]{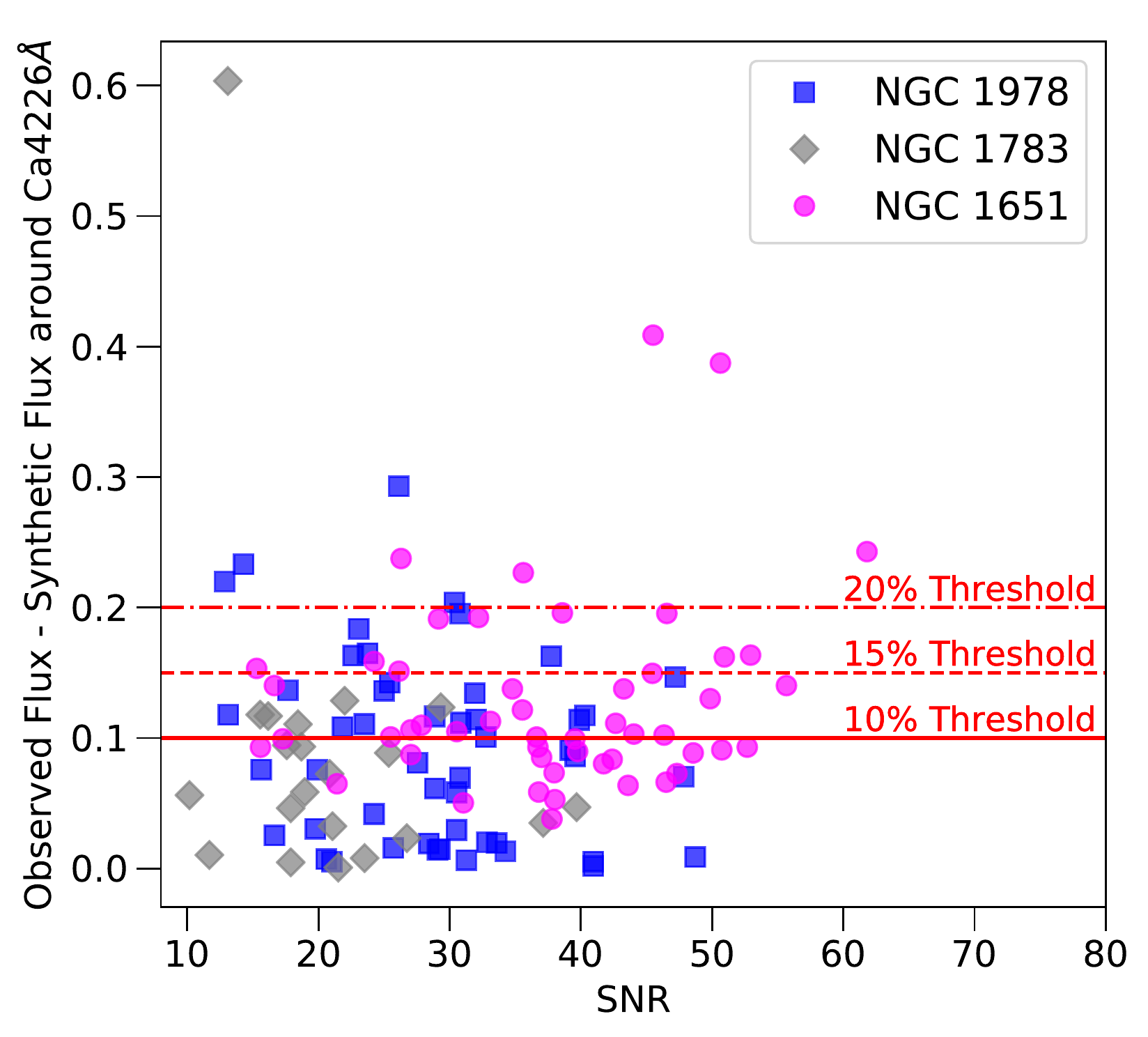}
    \caption{Difference in Ca(4226\AA) flux between the observed and the synthetic spectra for the stars in our samples. Different colours and markers indicate different cluster stars as reported in the legend. The horizontal red lines indicate the different thresholds used to determine membership of the stars to each cluster.}
    \label{fig:diff_ca}
\end{figure}

To quantify the difference between observed and synthetic spectrum around the two selected lines, we first locally normalised both the observed and synthetic spectra around the lines of interest (around the grey shaded areas in Fig. \ref{fig:mem_ca_fit}). 
Then, we determined the centroid of both observed and synthetic lines by performing a Gaussian fit, in the grey shaded area reported in Fig. \ref{fig:mem_ca_fit}. We finally calculated the difference in flux between the observed and synthetic line in the centroid\footnote{The Gaussian fit was performed to remove small shifts in wavelength between the line centroid of the observed spectrum and the synthetic one.}. We applied different thresholds for the membership selection, a more stringent 10\%, a 15\% and a 20\% of flux difference, for both Ca(4226\AA) and H$\gamma$. 
Choosing stars with a lower threshold does not imply choosing stars with better SNR. 
This can be seen in Figure \ref{fig:diff_ca}, which shows the difference in flux between the observed and synthetic spectrum around the Ca(4226\AA) line as a function of the total SNR of the spectra. The red lines indicate the different thresholds considered for the membership selection. The 10\% threshold includes at least half of the stars for NGC~1783 (grey diamonds) and NGC~1978 (blue squares), while only $\sim$40\% of the observed
stars are included for NGC~1651 (magenta circles). There is not much difference when changing threshold from 15\% to 20\%. 

The analysis was performed for both thresholds at 10\% and 15\%. However, we decided to consider the results for the more stringent and conservative threshold at 10\%, by checking visually the spectra and the goodness of fit one by one. 
Hence, throughout the paper we will report only the results for the threshold at 10\%, unless otherwise specified. 

Measured properties for all the analysed stars
are reported in Table \ref{tab:info_spectra}. We obtain 24 member stars for NGC~1978, 21 for NGC~1651 and 12 for NGC~1783. 

\begin{figure*}
\centering
\includegraphics[scale=0.53]{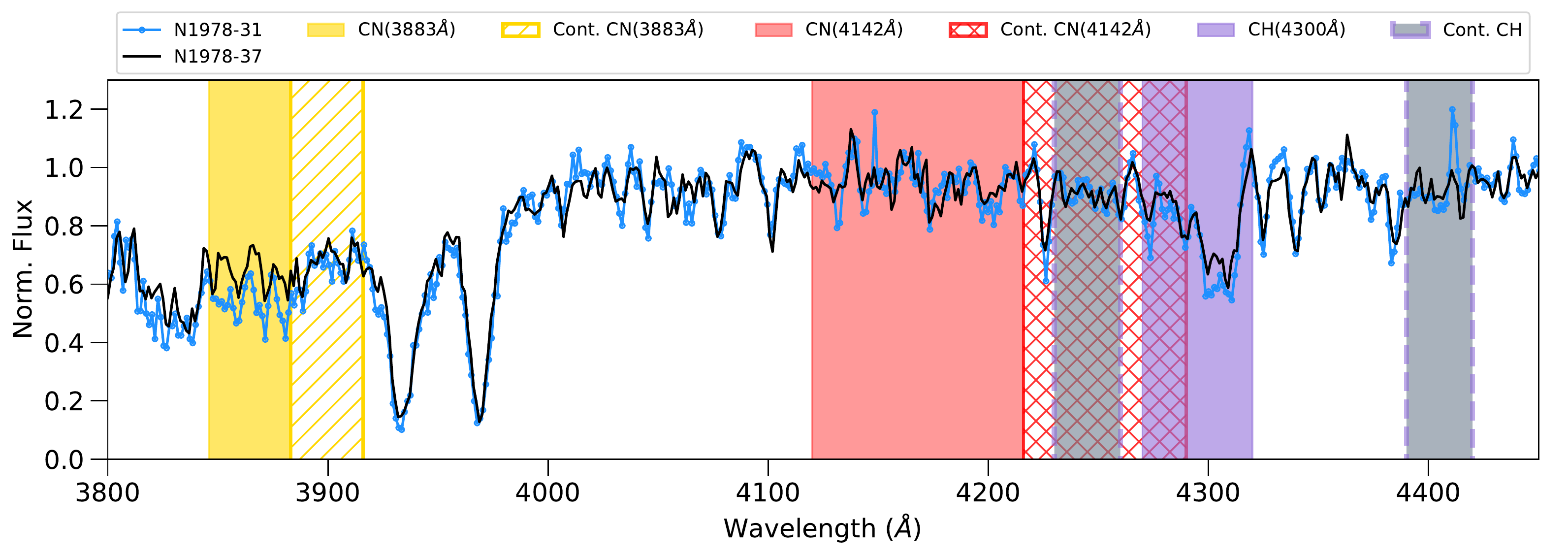}
\caption{CN-rich (blue) and CN-poor (black) spectra of the member stars N1978-31 and N1978-37, respectively. These have similar atmospheric parameters and SNR ($SNR_{31}=33$, $SNR_{37}=39$, \teff$_{,31}=4740$ K, log(g)$_{31}=2.6$, \teff$_{,37}=4970$ K, log(g)$_{37}=2.7$). 
The yellow, red and purple shaded areas indicate the spectral regions used to measure the two CN and the CH indices, respectively. The area hatched with yellow diagonal lines represents the continuum band for the CN(3883\AA), the area hatched with red crosses indicates the continuum for the CN(4142\AA) while the grey-shaded areas represent the two bands used for the continuum of the CH. 
} 
\label{fig:spectra_bands}
\end{figure*}

\subsection{Atmospheric Parameters Determination}\label{subsec:params}

Atmospheric parameters were obtained from photometry. The effective temperatures were calculated by using the $V-I$, $V-J$ and $V-Ks$ colour in the 
$T_{\rm eff}$-colour calibrations provided by \cite{ramirez05}. The final effective temperatures (\teff) were then computed by averaging the
three estimations. 
For the values of metallicity, distance modulus and extinction, we used BaSTI isochrones (\citealt{pietrinferni04}, \citealt{hidalgo18}). 
We started from values already reported in the literature and we slightly varied these to find isochrones that reproduced the shape of the cluster CMDs, see Fig. \ref{fig:cmds}. The final and used values are reported in Table \ref{tab:infolog}. 
Starting values of metallicities from high resolution spectroscopy are from \cite{mucciarelli08}
, while values of ages, distance moduli and extinction are from \cite{goudfrooij14} and \cite{martocchia18a}. 
Next, we calculated the surface gravity, i.e. log(g), through the Stefan-Boltzmann law, by using the previously derived $T_{\rm eff}$ and the distance moduli and stellar masses listed in Table \ref{tab:infolog}. We also used $M_{\odot}=1.989\times10^{33}$ g, $L_{\odot}=3.828\times10^{33}$ erg/s and $M_{\rm Bol, \odot}=4.75$ mag \citep{andersen99}. Bolometric corrections were computed based on the relations presented in \cite{alonso99}, using $V$ magnitudes. 
The adopted atmospheric parameters are listed in Table~\ref{tab:info_abu}.

\subsection{Index and Abundance Calculations}\label{subsec:ind}

CH and CN index measurements were 
calculated using the definitions by \cite{norris79,norris81}, as done in \cite{lardo13} and in our previous studies \citep{hollyhead17,hollyhead18,hollyhead19}. The definitions are the following:
\begin{equation}
CN (3883\angstrom) = -2.5 \log \bigg(\frac{\int_{3846}^{3883} F_{\lambda} d\lambda}{\int_{3883}^{3916} F_{\lambda} d\lambda}\bigg),
\label{eq:blue_cn}
\end{equation}
\begin{equation}
CN (4142\angstrom) = -2.5 \log \bigg(\frac{\int_{4120}^{4216} F_{\lambda} d\lambda}{\int_{4216}^{4290} F_{\lambda} d\lambda}\bigg),
\label{eq:red_cn}
\end{equation}
\begin{equation}
CH (4300\angstrom) = -2.5 \log \bigg(\frac{\int_{4270}^{4320} F_{\lambda} d\lambda }{1/2 \int_{4230}^{4260} F_{\lambda} d\lambda + 1/2 \int_{4390}^{4420} F_{\lambda} d\lambda} \bigg), 
\label{eq:ch}
\end{equation}
where $F_{\lambda}$ represents the measured intensity of the spectra at wavelength $\lambda$. 
Errors for the indices were estimated assuming Poisson statistics, as in \cite{vollmann06}.

Index measurements for all targets are listed in Table~\ref{tab:info_abu}.
Figure~\ref{fig:spectra_bands} shows the spectral windows used to calculate the band strength and the continuum of the indices. Superimposed are the spectra of two giants in NGC~1978.

Carbon and nitrogen abundances were computed by fitting observed spectra with synthetic templates around the CH and CN(3883\AA) indices, respectively. To calculate synthetic spectra, we adopted the atmospheric parameters derived in Sect. \ref{subsec:params}, along with the metallicities listed in Table \ref{tab:infolog}.

We assume a microturbulent velocity $v_t=2$ km/s for all the stars.
Both atomic and molecular line lists were taken from the
most recent Kurucz compilation from F. Castelli's website\footnote{\url{http://wwwuser.oats.inaf.it/castelli/linelists.html}}.
CH molecular lists are from \cite{masseron14}, while CN line lists are from \cite{brooke14}.
Model atmospheres were calculated with 
the ATLAS9 code \citep{castelli04} using the appropriate temperature and surface gravity for each star. 
A solar-scaled composition from \cite{asplund09} was assumed. Also, a solar carbon isotopic ratio has been used.

We generated model spectra using SYNTHE \citep{kurucz05} with a range of chemical abundances of 0.2 dex as step size.
Then, we fitted our observed spectra with a $\chi^2$ minimisation algorithm to find the model that best fit our data.  
Derived abundances are listed in Table \ref{tab:info_abu}. 

\begin{table}
\centering
    \caption{Values of the errors from different sources for the abundances of C and N in dex. See text for more details.}
    \label{tab:errors}
        \begin{tabular}{c|cc|cc|cc}
        & \multicolumn{2}{c|}{\textbf{NGC~1651}}&\multicolumn{2}{c|}{\textbf{NGC~1783}}&\multicolumn{2}{c}{\textbf{NGC~1978}}\\
        & {$\delta$C}&{$\delta$N}&{$\delta$C}&{$\delta$N}&{$\delta$C}&{$\delta$N}\\
        \hline
        $\delta$Teff & 0.11 & 0.24 & 0.10 & 0.24 & 0.11 & 0.24 \\
        \hline
        $\delta$log(g) & 0.03 & 0.05 & 0.02 & 0.05 & 0.03 & 0.05\\
        \hline
        $\delta v_t$ & 0.16 & 0.06 & 0.13 & 0.06 & 0.16 & 0.06\\
        \hline
        $\delta$C & - & 0.22 & - & 0.28 & - & 0.22\\
        \hline
        Tot. Systematic & 0.19  & 0.34 & 0.18 & 0.36 & 0.19 & 0.34\\
        \hline
        Statistical & 0.05 & 0.07 & 0.08 & 0.09 & 0.05 & 0.07\\
        \hline
        \textbf{Total Error} & 0.20 & 0.35 & 0.20 & 0.37 & 0.20 & 0.35 \\
        \hline
        \end{tabular}
\end{table}

To estimate the errors associated to the abundance measurements, we consider a typical error of $\pm$150 K in the effective temperature, an error of $\pm$0.2 dex in log(g) and an error of $\pm$1 km/s in the microturbulent velocity\footnote{As the goal of the paper is to study intrinsic spreads and not the absolute values of N and C abundances, the error on metallicity is not considered here, because we can assume that all stars in the same cluster have the same metallicity.}. 
An error analysis was performed by varying one atmospheric parameter at a time 
while keeping the others fixed and re-determining the abundances for the coldest and warmest stars in the sample (see e.g. \citealt{lardo13,lardo16,hollyhead18}). 
In Table~\ref{tab:errors}, the changes in abundances are given as a function of the stellar parameters \teff, $\log$(g),  microturbulent velocity $v_t$, and C for the N abundance error estimates. 
As it can be seen, the largest error from the adopted parameters come from the effective temperature on both the abundances. Also, the C abundances have relatively large uncertainties due to the errors on the 
microturbulence, with respect to N. This is expected, as the microturbulence mainly affects a relatively strong spectral feature such as CH, compared to the weaker CN band.
For the calculation of the [N/Fe] ratios, we used the previously derived [C/Fe] abundances, since the CN bands depend on both N and C abundances. In Table \ref{tab:errors}, the error on N which derives from the error on the C abundances is also reported. It is clear that this source of error is quite important and mainly determine the different uncertainties between C and N.

Due to the covariance of the atmospheric parameters, it is not correct to sum the single errors reported in Table \ref{tab:errors} in quadrature (see \citealt{mcwilliam95}). Hence, to calculate the total systematic uncertainties, we re-estimated the abundances by means of Monte Carlo (MC) simulations, where the parameters \teff, $\log$(g), $v_t$ (and the C abundance when estimating the error on N abundances) are left free to vary simultaneously. We generated 500 synthetic spectra with parameters drawn from normal distributions centred on the measured \teff, $\log$(g), $v_t$ and [C/Fe] (from Table \ref{tab:info_abu}) and with width equal to the respective errors reported above, for both the coldest and warmest star in the sample, for each cluster. We then calculated the standard deviation of the 500 simulated abundances for the warmest and coldest stars. The mean value between these two represents the total systematic error, which is also reported in Table \ref{tab:errors}.    

Finally, the statistical error associated to the measurements was estimated by means of additional MC simulations. To this aim, we generated 500 synthetic spectra with best fit parameters and injected them with Poissonian noise to reproduce the noise conditions observed around the molecular features. These uncertainties (reported in Table \ref{tab:errors}) are of the order of 0.05-0.09 dex, being larger for the lower SNR spectra of NGC~1783. 
The systematic and statistical errors were added in quadrature and gave the final errors reported in Table \ref{tab:errors} and \ref{tab:info_abu}.

\section{Results}\label{sec:res}

In this Section, we report the results from the indices and abundance calculations. The reason why we used both indices and abundances in our analysis is that indices are calculated directly from the spectra, while the abundances are estimated assuming certain stellar parameters such as effective temperature, gravity, microturbulence. These assumptions propagate and amplify the uncertainties. A spread might be observable in the CN indices but not in the N abundances, due to the larger errors.

\subsection{Indices}\label{subsec:res_ind}

\begin{figure}
\centering
\includegraphics[scale=0.55]{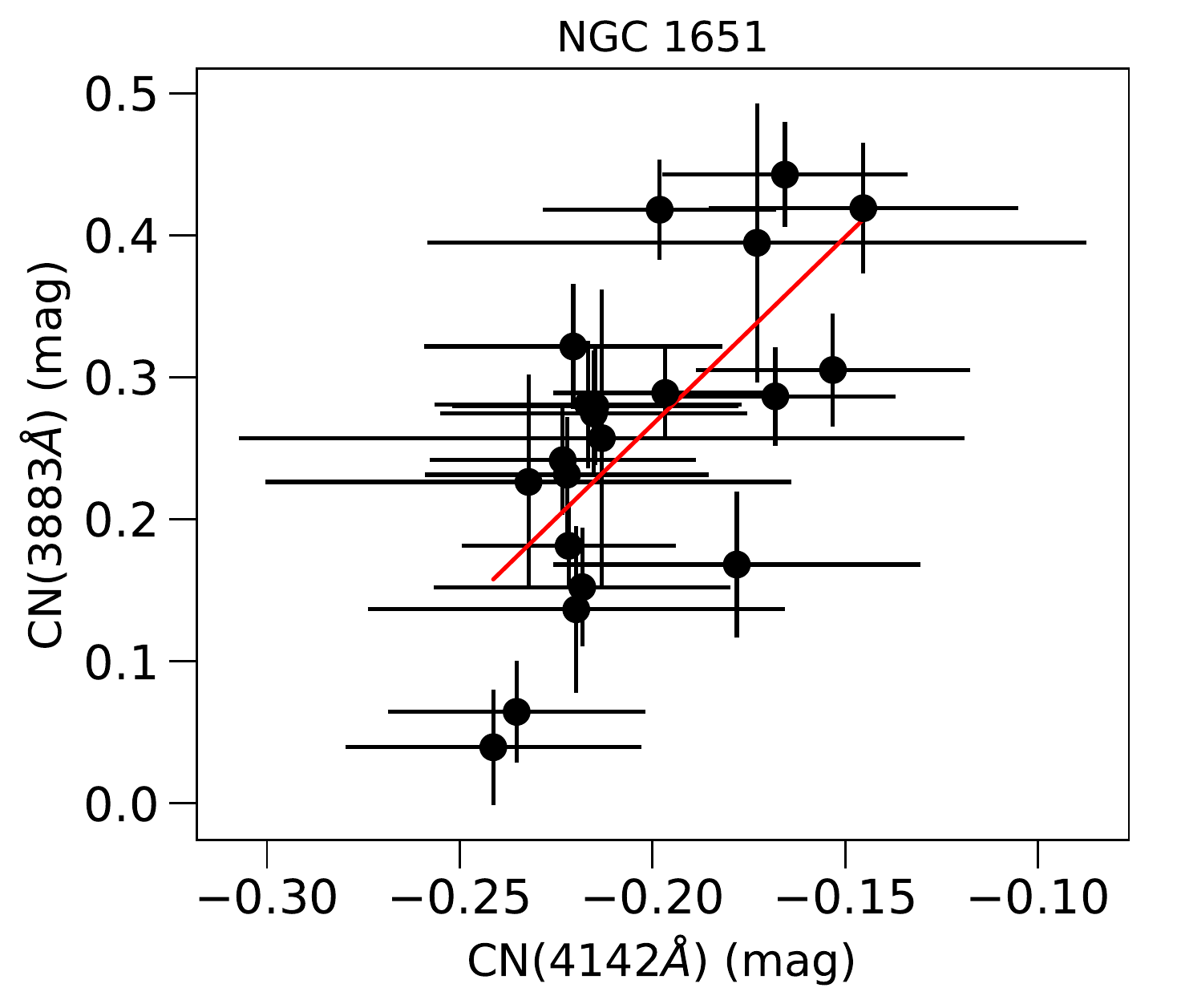}
\caption{CN at 4142\AA\, versus CN at 3883\AA\, for NGC~1651. The red solid line indicates a linear fit to the points.}
\label{fig:CN_corr}
\end{figure}

We measured two different indices for the CN: the one at 3883\AA\, and the one at 4142\AA. 
Figure \ref{fig:CN_corr} shows the comparison between the two CN absorption bands for NGC~1651.
The red solid line represents a linear fit to the data. The Spearman's rank correlation coefficient is $\rho_s=0.74$ and the probability that the two indices are not correlated is $\sim 10^{-4}$, indicating the presence of a strong positive correlation. The same is valid for the other two clusters.
Just note that the variations in the CN at 3883\AA\, are significantly larger than those measured for the CN at 4142\AA. From Fig. \ref{fig:CN_corr}, the former spans from $\sim$0 to $\sim$0.4 mag while the latter spans from $\sim -0.25$ to $-0.15$ mag. Hence, we decided to consider only the CN(3883\AA) for the rest of our analysis \citep{harbeck03,pancino10}.

\begin{figure*}
\centering
\includegraphics[scale=0.42]{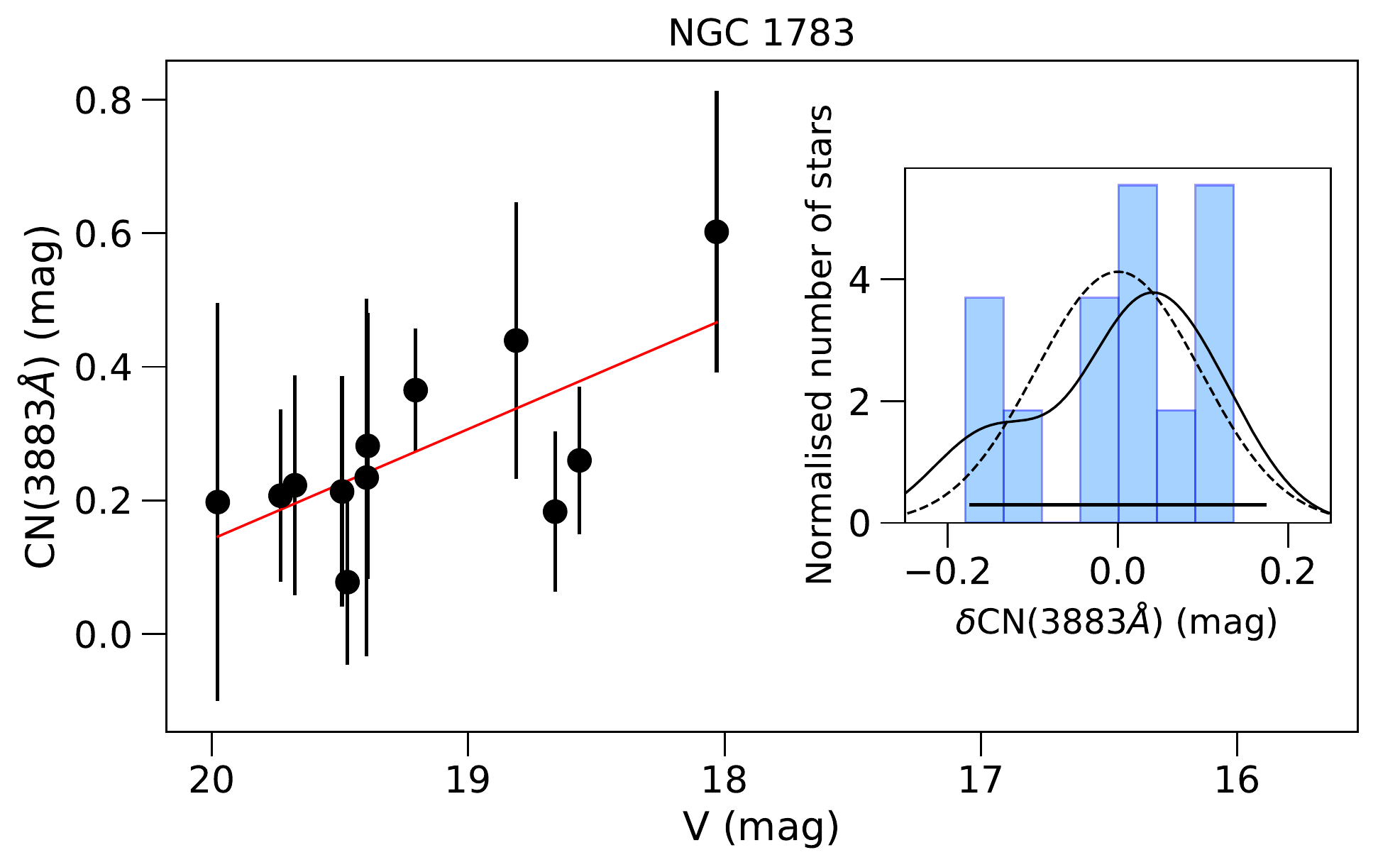}
\includegraphics[scale=0.42]{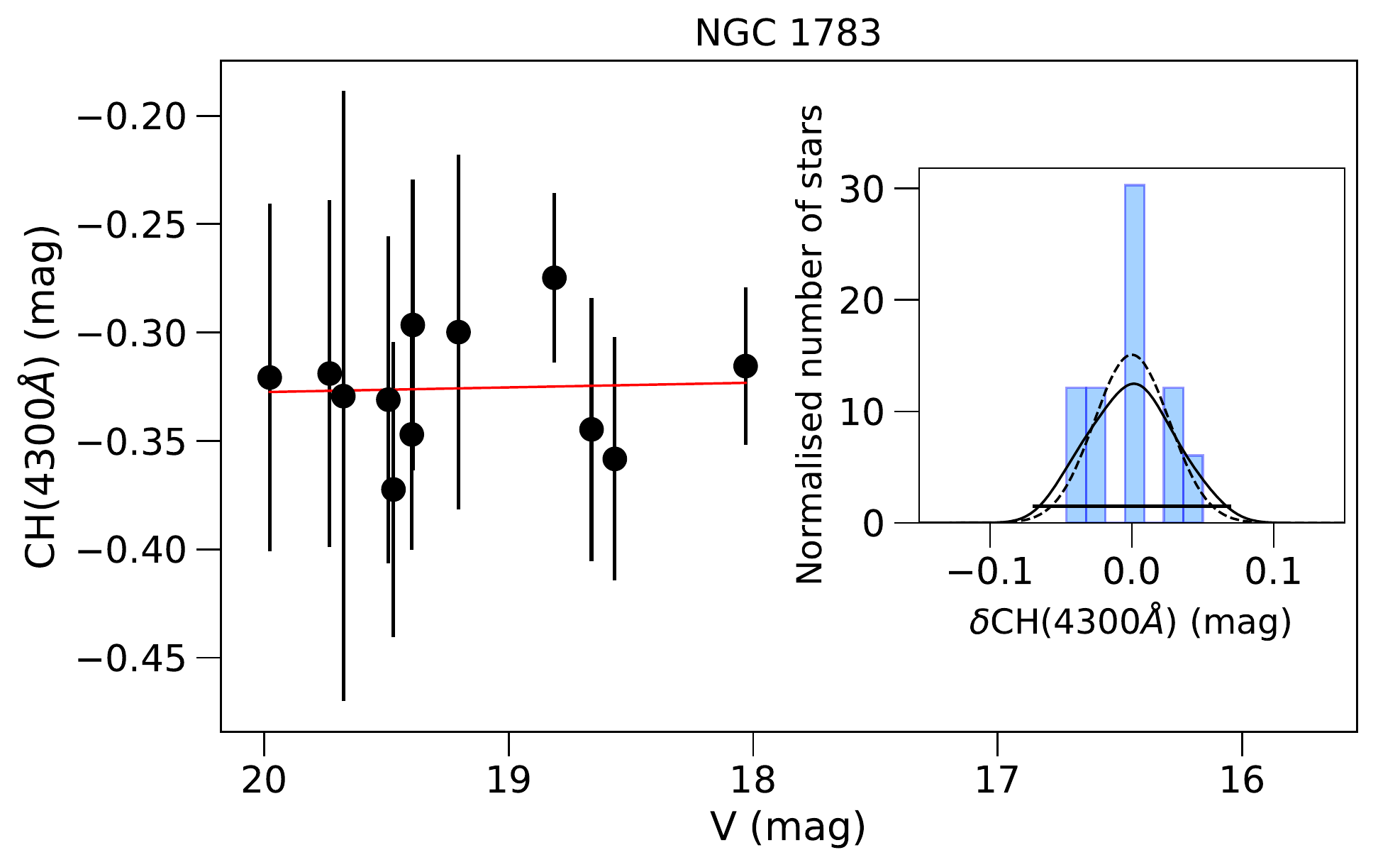}
\includegraphics[scale=0.42]{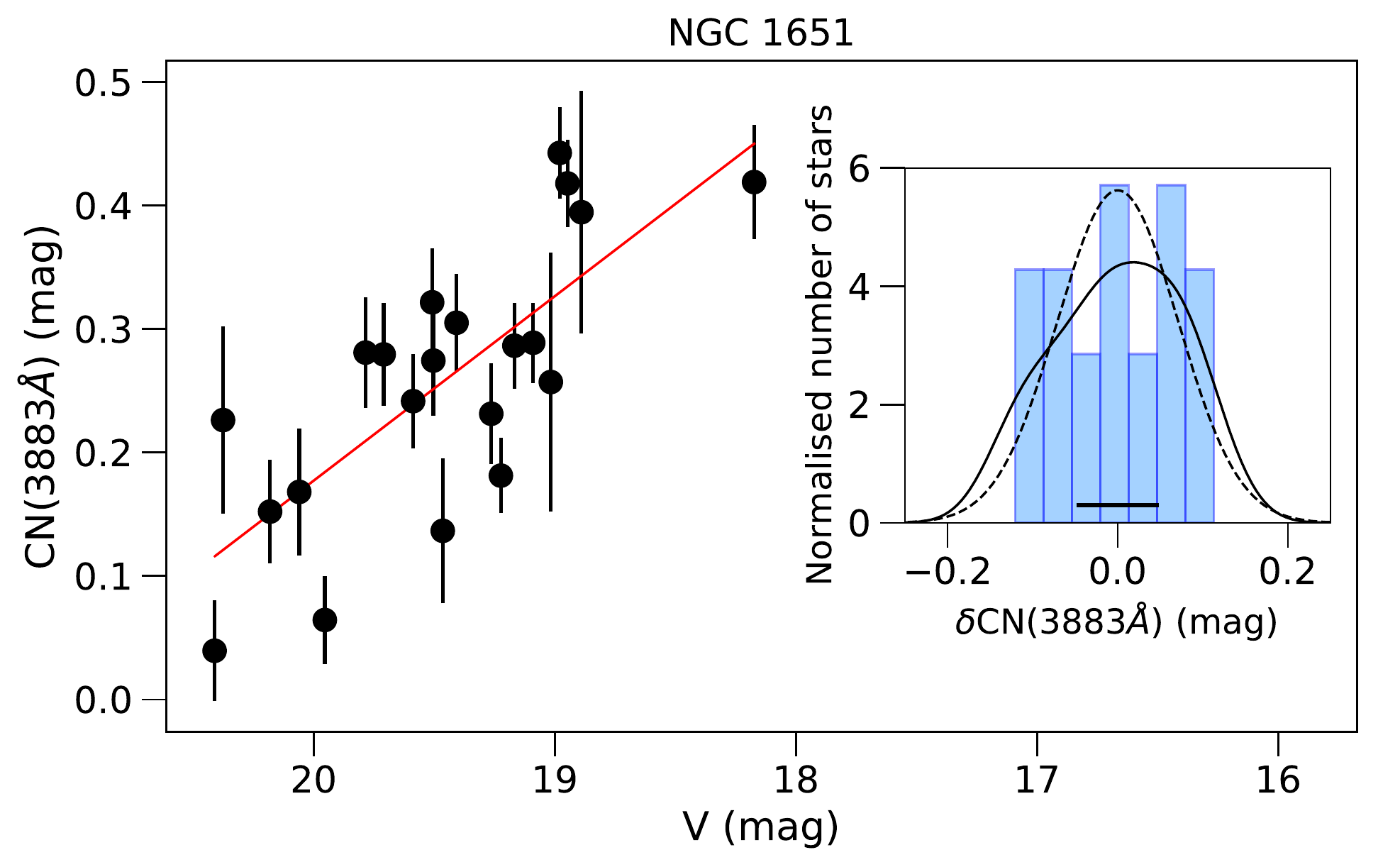}
\includegraphics[scale=0.42]{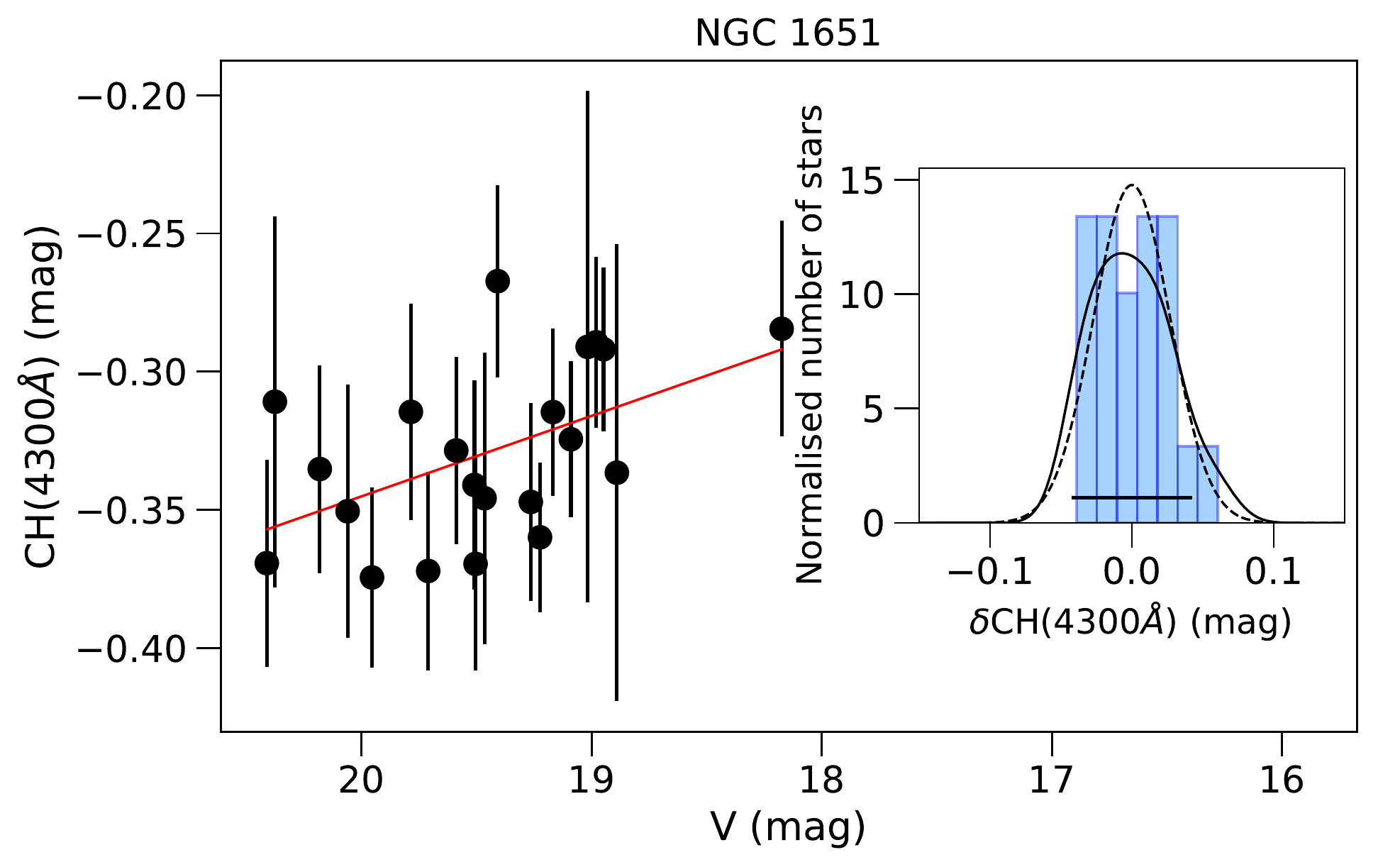}
\includegraphics[scale=0.42]{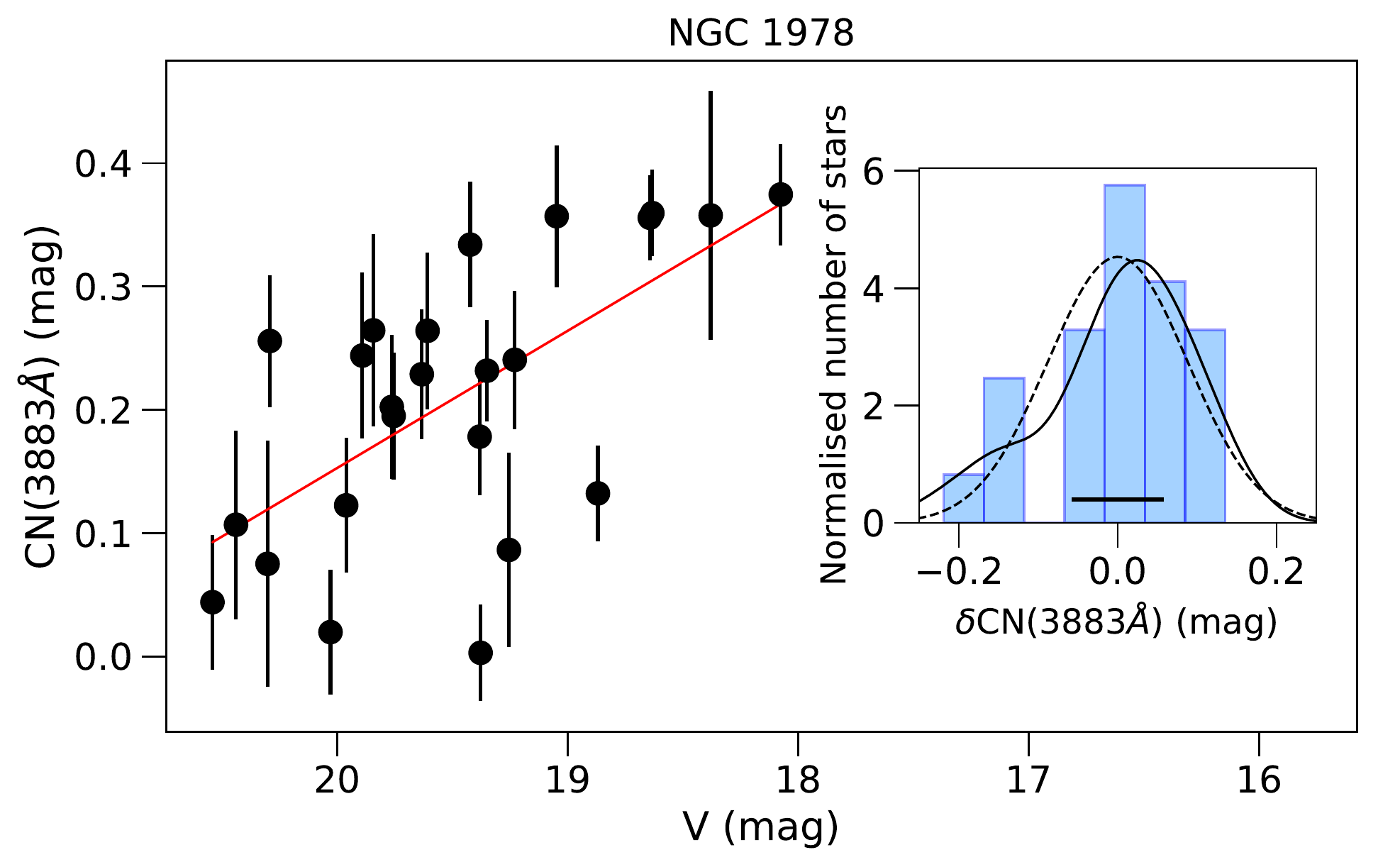}
\includegraphics[scale=0.42]{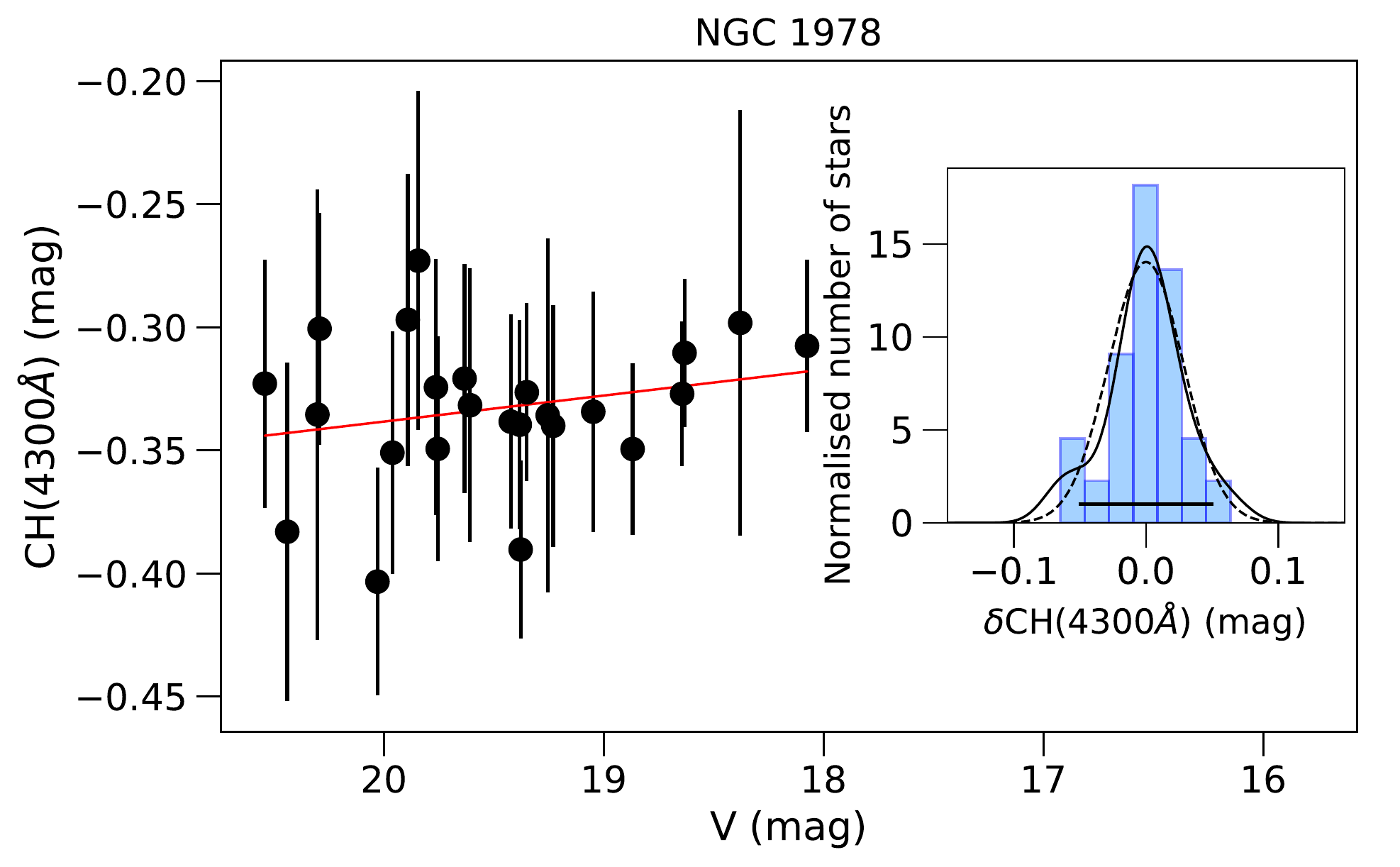}
\caption{From left to right: CN(3883\AA) and CH indices as a function of the V magnitude for cluster members, for NGC~1783 (upper panels), NGC~1651 (middle) and NGC~1978 (lower). The red lines show a linear fit to the data. In the inset we show the histograms of the residuals in each respective index with respect to the linear fit. The kernel density estimators are shown as black solid curves, while the best fit gaussian distributions are plotted as dashed black lines. Mean errors of residuals are plotted at the base of each histogram.}
\label{fig:CN_CH}
\end{figure*}

Fig. \ref{fig:CN_CH} shows the CN(3883\AA) (left), and CH (right) indices as a function of the $V$ magnitudes for the selected members of NGC~1783 (upper panels), NGC~1651 (middle panels) and NGC~1978 (lower panels); the red solid lines denote the linear fit between the quantities. For a more quantitative analysis, the insets in each Fig. show the histograms of the indices residuals $\delta$ (with respect to the linear fit) compared to kernel density estimator (KDE) distributions (black solid curves) and Gaussian distributions (black dashed curves).
Comparisons between the observed distributions and the associated uncertainties (shown at the base of the histograms) reveal that no spread is detected in the CH index, within the errors.

The left panels of Fig. \ref{fig:CN_CH} show that the CN index increases as a function of magnitude, hence it is important to remove the dependence on the luminosity to make a like-with-like comparison of all the stars within the sample. 
The histogram distributions in the insets of the left panels of Figures \ref{fig:CN_CH} show visually that there might be a spread within the rectified $\delta$CN values which is not consistent with the mean error.
Additionally, the KDEs distributions in NGC~1783 and NGC~1978 (solid black line) look different from the respective Gaussian fits (dashed black line), with a few stars with negative residuals that might indicate a spread. 

To quantitatively assess the presence of an intrinsic spread on the $\delta$CN, we adopted a maximum-likelihood method as reported in the Appendix D of \cite{kamann14}. We assumed that the intrinsic spread and the uncertainties are Gaussian, hence the probability to measure a certain $\delta$CN is:
\begin{equation}
    p(\delta CN_i) = \frac{1}{\sqrt{2\pi (\sigma_{\delta CN}^{2}+\epsilon_{i}^{2})}}\exp{\bigg[- \frac{(\delta CN_i - \overline{\delta CN})^2}{2(\sigma_{\delta CN}^{2}+\epsilon_{i}^{2})}\bigg]},
\label{eq:p1}
\end{equation} 
where $\epsilon_i$ represents the uncertainty on the measured $\delta CN$ of the star $i$, where $i\in[1:N]$ and N represents the total number of observed stars. Hence, the likelihood of observing the data is the product of all the individual $i$ probabilities. We then minimised the negative log-likelihood with a Markov Chain Monte Carlo (MCMC, \citealt{mcmc}) code to obtain the value of the intrinsic spread ($\sigma_{\delta CN}$) within each cluster.

Interestingly, we found that the intrinsic spread in NGC~1783 is consistent with zero, while a spread in CN is detected for NGC~1978 and NGC~1651, being $\sigma_{\delta CN}$(NGC~1978)$=0.08^{+0.02}_{-0.01}$ mag and $\sigma_{\delta CN}$(NGC~1651)$=0.06^{+0.02}_{-0.01}$ mag. The  errors are calculated on the 16th and 84th percentile of the MCMC distributions. 
Figure \ref{fig:spectra_bands} shows the comparison between a CN-poor (black) and CN-rich (blue) member star of NGC~1978 that have similar atmospheric parameters, i.e. effective temperature and gravity. Indeed, it is possible to see the difference around the CN at 3883\AA\, (yellow shaded area). The same is also observed in NGC~1651. 
We also ran the code when relaxing the membership threshold to 15\%, and we kept finding that the intrinsic spread in NGC~1783 is consistent with zero, while it is still 
significant for NGC~1978 and NGC~1651.

\begin{figure}
\centering
\includegraphics[scale=0.44]{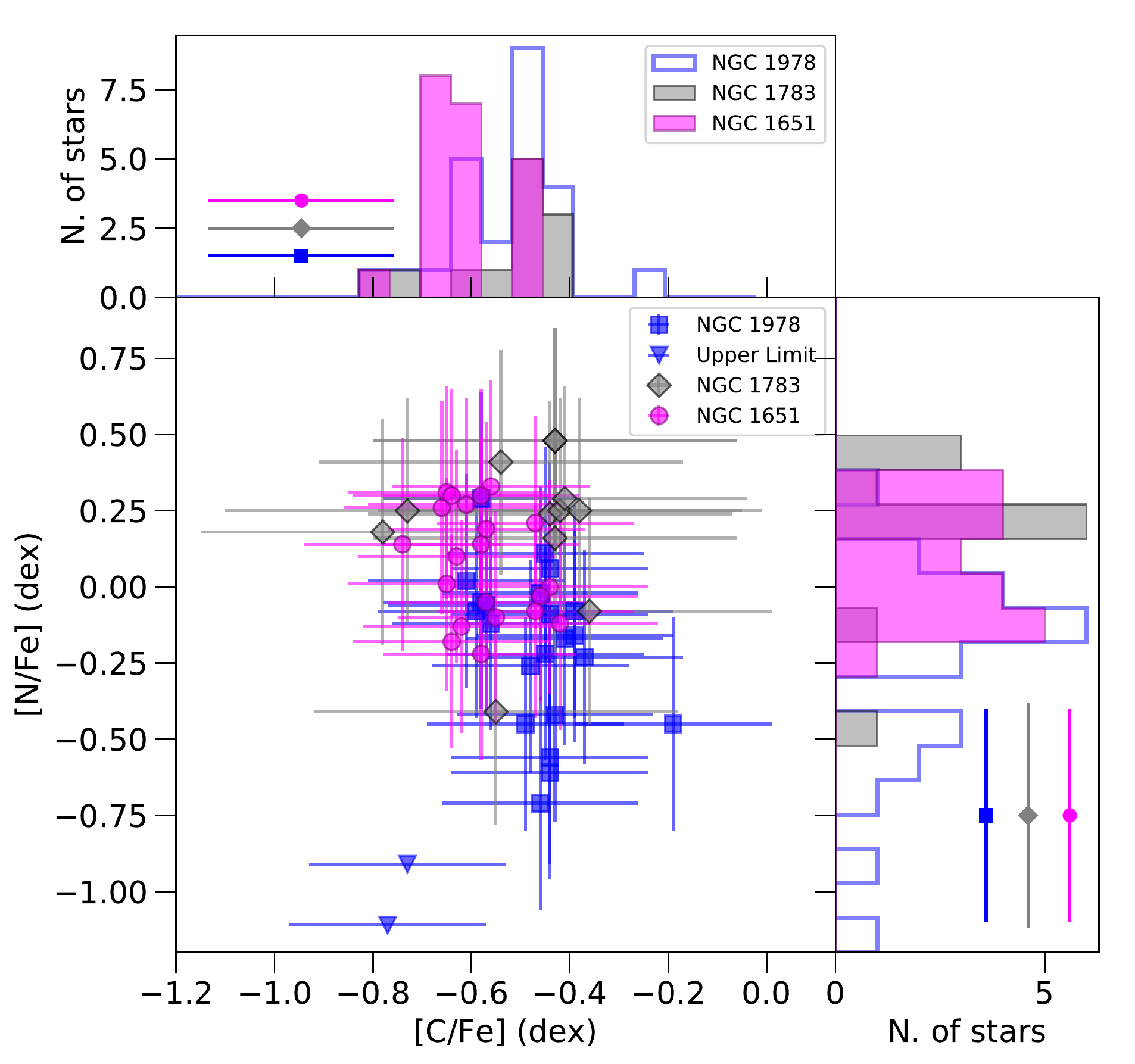}
\caption{[C/Fe] abundances versus [N/Fe]. Different colours and markers indicate different clusters as reported in the legend. On the upper and side panels, the histograms of the distribution of the [C/Fe] and [N/Fe] abundances are reported, respectively, along with the mean error of the sample for each cluster.}
\label{fig:comparison_abu}
\end{figure}

We also fit the $\delta$CN data with Gaussian Mixture Models (GMMs) to identify the presence of multiple Gaussian components in the distributions of each cluster, i.e.\ two or more populations with different N abundances. This was done by using the GMM code by \cite{muratovgmm}. The code found that a bimodal population is preferred for NGC~1978 but not for NGC~1651. However, the significance of the bimodality for NGC~1978 is very low, according to the GMM p-value.

\subsection{Abundances}\label{subsec:res_abu}
Next, we analysed the C and N abundances with the aim to translate the observed spreads in CN into an internal N variation. 
First, we checked that there was no dependence on magnitude in the calculated abundances.
Figure \ref{fig:comparison_abu} shows [C/Fe] versus [N/Fe] abundances for all the three clusters. Blue squares represent NGC~1978 member stars, magenta circles indicate NGC~1651 and grey diamonds indicate NGC~1783. On the top and right panels, the histograms of the distributions in [C/Fe] and [N/Fe] are reported along with errors on the abundances. No spread is obvious in [C/Fe] abundances. 
Regarding the [N/Fe] abundances, there is also no visible strong anti-correlation or spread. 
Visually for [N/Fe], there is a larger spread in NGC~1978 than in the other two clusters.

To quantify this, we also calculated the intrinsic spread on N with the same maximum likelihood approach that was exploited in Section \ref{subsec:res_ind}. We obtained that the intrinsic spread in N is consistent with zero for all the clusters.
Hence, for NGC~1651 and NGC~1978, where a CN spread is observed, no significant N intrinsic spread can be constrained from abundances \textit{alone}, most likely due to the large uncertainties on the abundance calculation (see Sect. \ref{subsec:ind}). 
However, we are confident that the intrinsic variations in CN are present, which indicates solid evidence for the presence of MPs within these two clusters. 

\begin{figure}
\centering
\includegraphics[scale=0.44]{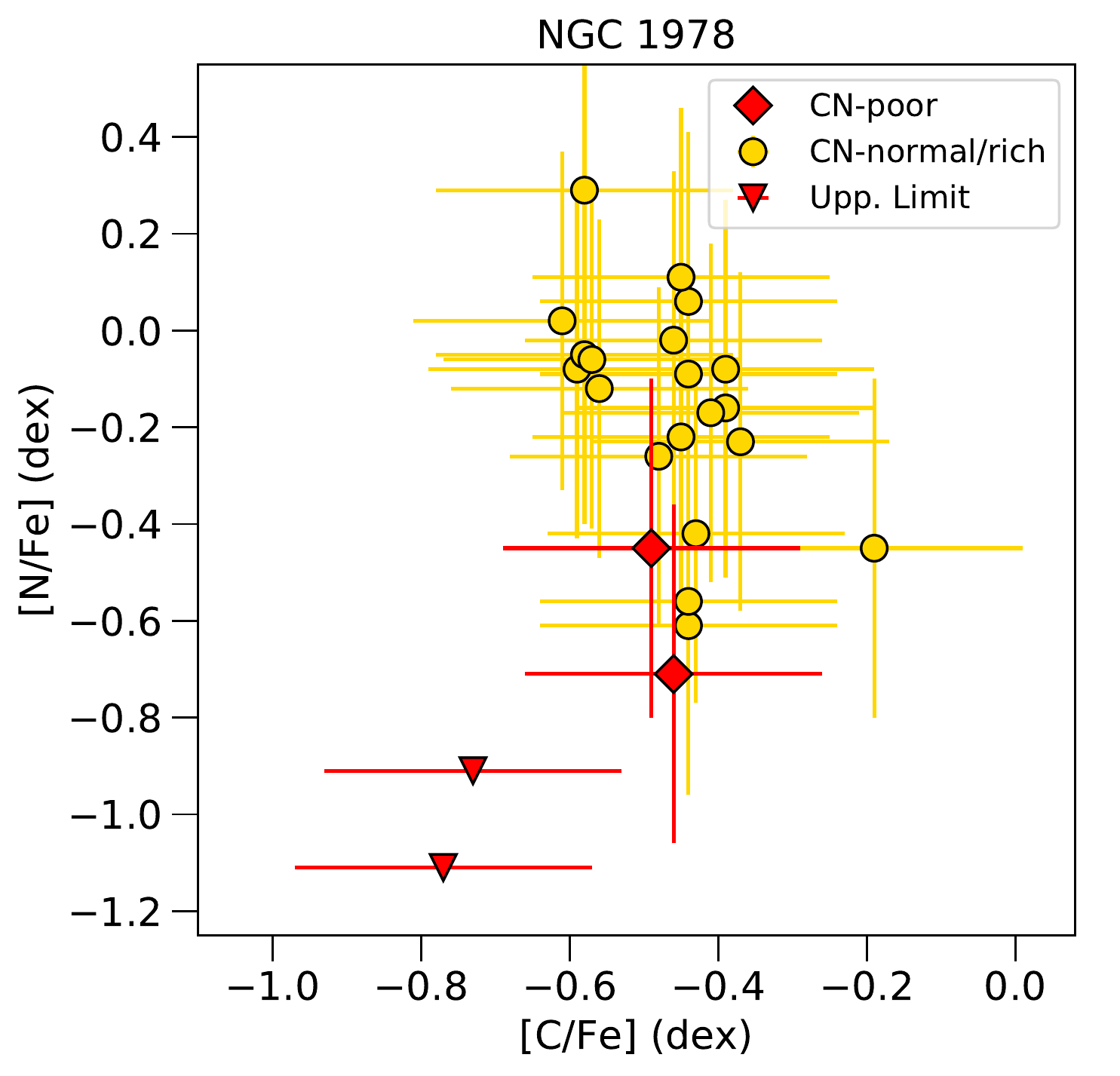}
\caption{[C/Fe] vs [N/Fe] abundances for NGC~1978 member stars. The red filled diamonds (and upside triangles for upper limits in [N/Fe]) indicate stars that have a $\delta$CN $< -0.1$ mag. See text for more details.}
\label{fig:mps_n1978}
\end{figure}

Through the maximum likelihood method applied on abundances, we were then only able to put an upper limit on the N spread of NGC~1783 $\Delta$[N/Fe]$\leq 0.4$ dex, and a $\Delta$[N/Fe]$\leq 0.2$ dex for NGC~1651,
at 2$\sigma$ confidence level.
For NGC~1978, as we observe an intrinsic spread in CN and a bimodality in the CN distribution is detected from the GMM fitting, we separated the two populations in the $\delta CN$ distribution to see where they lie in the [C/Fe] vs. [N/Fe] plane. 
We selected stars with $\delta$CN$<-0.1$ (see histogram in Fig. \ref{fig:CN_CH}), as this also corresponds to the minimum of the GMM distribution fit, i.e. where the two gaussian components cross. 
We show these CN-poor stars as red filled diamonds in Fig. \ref{fig:mps_n1978}, while yellow filled circles represent CN-normal/rich stars. We obtained a mean difference in N between the two populations of $\Delta$[N/Fe]$=0.63\pm0.49$ dex\footnote{As a test, we also checked this when relaxing the membership threshold to 15\% and we obtained a similar result,  $\Delta$[N/Fe]$=0.67\pm0.49$ dex.}, at $\sim 1.3 \sigma$ confidence level. The error is obtained by summing in quadrature the mean error on each population. As expected, CN-poor stars also show low [N/Fe] ratios, although these are scattered due to the large errors in the abundance estimation (see Table \ref{tab:errors}).

\section{Discussion and Conclusions}\label{sec:disc}

In this paper, we presented a spectroscopic study of RGB stars in three intermediate age ($\sim 1.7-2.3$ Gyr old) massive star clusters in the LMC, namely NGC~1783, NGC~1651 and NGC~1978. High resolution spectroscopic studies of the same clusters were already carried out by \cite{mucciarelli08}. 
Interestingly,
no star-to-star variations in Na and O is significant within the clusters, although with rather large uncertainties. Na variations of the order of 0.07$\pm$0.01 dex were instead recently detected in NGC~1978 by \cite{saracino20b}.
In this work, we focused on two other elements, C and N. We used the ESO/VLT FORS2 low resolution multi-object spectrograph 
to look for intrinsic spreads in N abundances indicative for the presence of multiple populations within the clusters.
In particular, we measured indices for CH and CN, as well as C and N abundances for 24, 21 and 12 members in NGC~1978, NGC~1651 and NGC~1783, respectively. 

In all three clusters, we found no significant spread in CH or [C/Fe]. 
We found a statistically significant signal for the presence of star-to-star CN variations in NGC~1978 and NGC~1651. This is a strong indication that chemical anomalies are present within these two clusters. 

We were not able to quantify an intrinsic spread in [N/Fe] from the current abundance dataset (see Sect. \ref{subsec:res_abu}), possibly due to the large uncertainties. 
We put an upper limit on the N abundance of NGC~1651 $\Delta$[N/Fe]$\leq 0.2$ dex.
For NGC~1978, we instead quantified the internal N variation by 
separating the two populations from the CN indices, as the GMM fit reports a detection of bimodality in the $\delta$CN distribution (Sect. \ref{subsec:res_ind} and \ref{subsec:res_abu}). We obtained $\Delta$[N/Fe]$=0.63\pm0.49$ dex for NGC~1978, at $\sim 1.3 \sigma$ confidence level. 
For NGC~1783, we did not find evidence of both CN and N spreads, hence either MPs are absent in this cluster or they cannot be detected within the measurement errors. We put an upper limit $\Delta$[N/Fe]$\leq 0.4$ dex for NGC~1783. 

Both the indices and abundances analyses presented in this work clearly state that MPs in young clusters (when present) are small, with intrinsic spreads $\Delta$[N/Fe] of the order of 0.2-0.6 dex. 
However, we notice that at these magnitudes, our stars are undergoing the effect of the first dredge-up (FDU, \citealt{salaris20}). 
During the FDU, matter processed by H-burning reactions in the stellar interiors is dredged to the surface, due to the increasing depth of the convective envelopes, which reach layers where the abundance of N has attained  the equilibrium value of the CN cycle. This equilibrium abundance is higher than the corresponding initial solar scaled one, and this causes an increase of the surface N.
Such an increase after the FDU depends on the initial value of N. Higher initial N causes a lower increase at the FDU completion, because the nitrogen equilibrium abundance becomes comparable to the initial value,  hence the effect of the dredge up on the surface N is reduced.
At the end of the dredge-up, the N variation between two populations is expected to be smaller than the initial one. The N variations reported here should then be considered as a lower limit to the initial N spread between the two populations in the clusters. For NGC~1783, where the CN spread is not detected, it could also be that a N variation is too small to be detected due to the FDU effect.

The results reported here are consistent with the photometric work presented in \cite{martocchia18a}. By comparing the width of the RGB with isochrones with different chemical mixtures, we photometrically obtained a spread $\Delta$[N/Fe]$\sim+0.5$ dex for NGC~1978 and put an upper limit on NGC~1783 of $\Delta$[N/Fe]$<+0.3$ dex \citep{martocchia18a}. Indeed, we observe a spread in the older clusters, namely NGC~1978 and NGC~1651, but none in the younger one, NGC~1783. NGC~1978 is around $\gtrsim2$ Gyr old (\citealt{mucciarelli07,martocchia18a}), while NGC~1651 is slightly younger ($\sim$2 Gyr, \citealt{goudfrooij14}) and NGC~1783 is younger than the two ($\lesssim$1.7 Gyr, \citealt{mucciarelli07b,goudfrooij11}). This is also observable in Fig. \ref{fig:cmds}, just by looking at the turn offs of each cluster: NGC~1783 has a turn off around $V\sim$20.2 mag, while this is $V\sim$20.6 mag for NGC~1651 and $V\gtrsim$21 mag for NGC~1978.

Additionally, in \cite{martocchia19}, we reported the presence of a correlation between the N spread and the age of the cluster, where ancient GCs are found to have larger N variations with respect to the younger ones.
As shown recently by \cite{salaris20}, the $\Delta$[N/Fe]-age correlation is affected by the FDU. 
As mentioned above, the corresponding change of surface N abundance depends on the initial N abundance but also on the mass of the star, hence its age \citep{salaris15,salaris20}.
The variation of the surface N with respect to the initial N abundance increases with increasing RGB stellar mass (decreasing
age of the population) and increasing metallicity.
The FDU is therefore one of the reasons why we do not expect large N spreads within such young clusters compared to ancient GCs, as we are not probing initial N abundances, but rather abundances 
modified through the FDU. 
However, \cite{salaris20} also showed that the FDU cannot entirely explain the $\Delta$[N/Fe]-age correlation. 

For future studies, it would then be critical to estimate the initial N spreads, i.e. not affected by the FDU and evolutionary effects, as a function of cluster age. This means targeting main sequence stars, where the FDU mixing has still not occurred. A pilot study has recently been performed by \cite{cabrera20}, where they looked for abundance variations in the MS of the $\sim$1.5 Gyr old, massive ($\sim10^5$\msun) cluster NGC 419, making a comparison with Galactic GCs such as 47 Tuc, NGC 6352 and NGC 6637 which have similar metallicities ([Fe/H]$=-0.7$ dex). 
By using HST photometry to analyse MS stars that have the same range in stellar masses of stars where MPs are found in old GCs, i.e. $\sim 0.75-1.05$\msun, they found that the colour distributions of NGC 419 in the lower MS are consistent with what is expected from a cluster with homogeneous abundances. However, the sensitivity of the current dataset cannot exclude small initial abundance variations. It is thus necessary to probe a dependency on mass and age by expanding the sample to other clusters. The HST or the upcoming James Webb Space Telescope will be the facilities necessary for such future follow-up studies. 
Spectroscopically, a huge collecting power as well as high spatial resolution to probe dense cluster centres will be required to observe such faint stars, hence this will be only possible in the era of the extremely large telescopes. Establishing a $\Delta$[N/Fe] vs age spectroscopic correlation when initial N spreads are considered will be an extremely useful constraint for any model aimed at explaining the origin of MPs.

\section*{Acknowledgements}
We thank the referee Chris Sneden for the positive and constructive report that helped strengthen the paper.
We gratefully acknowledge Maria-Rosa Cioni and Florian Niederhofer for providing us with the VMC photometric catalogues and we thank Alessio Mucciarelli for useful help and discussions. 
SM, SK and NB gratefully acknowledge financial support from the European Research Council (ERC-CoG-646928, Multi-Pop). NB also acknowledges support from the Royal Society (University Research Fellowship).
This study was enabled by a Radboud Excellence fellowship from Radboud University in Nijmegen, the Netherlands. CL acknowledges funding from  Ministero dell'Università e della Ricerca through the Programme ``Rita Levi Montalcini'' (grant PGR18YRML1). 
ED acknowledges financial support 
from the project {\it Light-on-Dark} granted by MIUR through
PRIN2017-2017K7REXT.

\section*{Data Availability}

The data underlying this article are available in the article and in its online supplementary material.



\bibliographystyle{mnras}
\bibliography{fors2} 


\begin{landscape}
\begin{table}
\centering
\caption{Measured stellar properties for the stars considered in our analysis. The full Table will be available in the online version of the paper$^1$.}
\begin{tabular}{l c c c c c c c c c c c c c c c c c c c c c c c c}
\hline
ID & R.A. & Dec. & $V$ & $I$ & $Y$ & $J$ & $Ks$ & RV & SNR & EW(Ca4226) & EW(CaHK) & Diff(Ca4226)$^2$ & Diff(H$\gamma$)$^2$  \\
 & (deg) & (deg) & (mag) & (mag) & (mag) & (mag) & (mag)  & (km/s) &  & (\AA) & (\AA) &  & \\
\hline
N1783-1 & 74.7434768 & -65.9865264 & 17.99 & 16.95 & 16.46 & 16.08 & 15.33 & 268.2 & 16 & 2.43$\pm$0.62 & 27.87$\pm$3.25 & 0.12 & 0.06\\
N1783-2 & 74.8583679 & -65.9918136 & 18.03 & 17.00 & 16.49 & 16.13 & 15.41 & 264.4 & 40 & 2.26$\pm$0.25 & 29.23$\pm$1.29 & 0.05 & 0.07\\
N1783-3 & 74.7812347 & -66.0178070 & 18.57 & 17.61 & 17.14 & 16.80 & 16.15 & 280.9 & 25 & 1.82$\pm$0.41 & 24.46$\pm$2.16 & 0.09 & 0.06\\
$\dots$ & $\dots$ & $\dots$& $\dots$ & $\dots$ & $\dots$& $\dots$ & $\dots$ & $\dots$& $\dots$ & $\dots$ & $\dots$& $\dots$ & $\dots$ \\
\hline
N1651-1 & 69.4243755 & -70.5277029 & 18.17 & 16.98 & 16.32 & 15.91 & 15.15 & 260.2 & 37 & 2.55$\pm$0.26 & 28.15$\pm$1.39 & 0.08 & 0.07\\
N1651-2 & 69.3138447 & -70.5614093 & 19.21 & 18.21 & 17.71 & 17.43 & 16.77 & 209.6 & 25 & 1.79$\pm$0.42 & 24.65$\pm$2.21 & 0.10 & 0.03\\
N1651-3 & 69.4244849 & -70.5645077 & 19.17 & 18.17 & 17.64 & 17.32 & 16.72 & 220.2 & 47 & 1.75$\pm$0.22 & 25.37$\pm$1.17 & 0.07 & 0.05\\
$\dots$ & $\dots$ & $\dots$& $\dots$ & $\dots$ & $\dots$& $\dots$ & $\dots$ & $\dots$& $\dots$ & $\dots$ & $\dots$& $\dots$ & $\dots$ \\
\hline
N1978-1 & 82.1954232 & -66.2018909 & 20.30 & 19.33 & 18.93 & 18.62 & 18.04 & 280.3 & 16 & 1.03$\pm$0.75 & 25.71$\pm$3.40 & 0.08 & 0.002\\
N1978-2 & 82.1969944 & -66.2043961 & 18.81 & 17.67 & 17.15 & 16.79 & 16.07 & 298.5 & 38 & 2.07$\pm$0.27 & 27.33$\pm$1.42 & 0.16 & 0.035\\
N1978-3 & 82.2085924 & -66.2077305 & 19.35 & 18.30 & 17.88 & 17.54 & 16.89 & 296.1 & 39 & 1.98$\pm$0.26 & 27.88$\pm$1.32 & 0.08 & 0.02\\
$\dots$ & $\dots$ & $\dots$& $\dots$ & $\dots$ & $\dots$& $\dots$ & $\dots$ & $\dots$& $\dots$ & $\dots$ & $\dots$& $\dots$ & $\dots$ \\
\hline
\label{tab:info_spectra}
\end{tabular}
\\
$^1$Note that the atmospheric parameters, the CN and CH indices and the abundances are reported in Table \ref{tab:info_abu}.
$^2$Member stars are selected having Diff(Ca4226)$<0.1$, and Diff(H$\gamma$)$<0.1$.
\end{table}

\begin{table}
\centering
\caption{Table of atmospheric parameters, indices and abundances for the stars considered in our analysis. The full Table will be available in the online version of the paper.}
\begin{tabular}{l c c c c c c c}
\hline
ID & \teff & log(g) & CN(3883\AA) & CN(4142\AA)& CH & [C/Fe] & [N/Fe] \\
 & (K) & (dex) & (mag) & (mag) & (mag) & (dex) & (dex)\\
\hline
N1783-1 & 4553 & 1.95 & 0.297$\pm$0.108 & -0.134$\pm$0.092 & -0.314$\pm$0.089 & -0.56$\pm$0.20 & 0.02$\pm$0.37\\
N1783-2 & 4575 & 1.98 & 0.602$\pm$0.210 & -0.139$\pm$0.037 & -0.315$\pm$0.036 & -0.43$\pm$0.20 & 0.48$\pm$0.37\\
N1783-3 & 4742 & 2.29 & 0.260$\pm$0.110 & -0.210$\pm$0.058 & -0.358$\pm$0.056 & -0.78$\pm$0.20 & 0.18$\pm$0.37\\
$\dots$ & $\dots$ & $\dots$ & $\dots$ & $\dots$ & $\dots$ & $\dots$ & $\dots$\\
\hline
N1651-1 & 4384 & 1.90 & 0.419$\pm$0.046 & -0.145$\pm$0.040 & -0.284$\pm$0.039 & -0.47$\pm$0.20 & -0.08$\pm$0.35\\
N1651-2 & 4856 & 2.59 & 0.162$\pm$0.062 & -0.234$\pm$0.057 & -0.331$\pm$0.056 & -0.67$\pm$0.20 & -0.11$\pm$0.35\\
N1651-3 & 4822 & 2.56 & 0.286$\pm$0.035 & -0.168$\pm$0.0312 & -0.315$\pm$0.030 & -0.58$\pm$0.20 & 0.14$\pm$0.35\\
$\dots$ & $\dots$ & $\dots$ & $\dots$ & $\dots$ & $\dots$ & $\dots$ & $\dots$\\
\hline
N1978-1 & 4961 & 3.05 & 0.075$\pm$0.100 & -0.252$\pm$0.093 & -0.335$\pm$0.091 & -0.59$\pm$0.20 & -0.08$\pm$0.35\\
N1978-2 & 4539 & 2.22 & 0.300$\pm$0.044 & -0.187$\pm$0.039 & -0.304$\pm$0.038 & -0.57$\pm$0.20 & -0.07$\pm$0.35\\
N1978-3 & 4769 & 2.57 & 0.232$\pm$0.041 & -0.226$\pm$0.037 & -0.326$\pm$0.033 & -0.46$\pm$0.20 & -0.02$\pm$0.35\\
$\dots$ & $\dots$ & $\dots$ & $\dots$ & $\dots$ & $\dots$ & $\dots$ & $\dots$\\
\hline
\label{tab:info_abu}
\end{tabular}
\end{table}

\end{landscape}


\appendix
\section{Radial Velocities Calculation}\label{sec:rv}

For the indices and abundance analysis, we need spectra shifted to the reference/laboratory wavelength scale. The shifts are primarily due to radial velocities (RVs) intrinsic to each star that include heliocentric velocity correction. We also measured additional systematic offsets due to the instrumental and observational setup.
 
We first estimated the RVs for each star in each exposure using the task \textit{fxcor} in \texttt{IRAF} with an appropriate synthetic spectrum as a stellar template for each cluster (see Sect.~\ref{subsec:ind} for the description of the synthetic spectra). 


The majority of the observations were carried out under good seeing ($<1"$) conditions. Hence, if the stars are not centred in the slit, this will generate a shift in the wavelength calibration and successively on the estimation of the RVs. 
To calculate the RV shift due to this effect, we followed the same approach as in \cite{harriszaritsky06}, \cite{kacharov17} and \cite{taibi18}.
We used the FORS2 through-slit images that are obtained before each science exposure to calculate the difference between the centroid of each star and and the centre of the 
slit in pixels. This was done for each exposure and for each star. The spatial shifts were then reported to velocity offsets according to the formula in Sect. 2.5 from \cite{harriszaritsky06} and subtracted to the previously obtained RVs. 
Finally, we calculated the heliocentric velocity with the \texttt{IRAF} task \texttt{rvcorrect} and added this offset to the RVs. 

\begin{figure}
\centering
\includegraphics[scale=0.46]{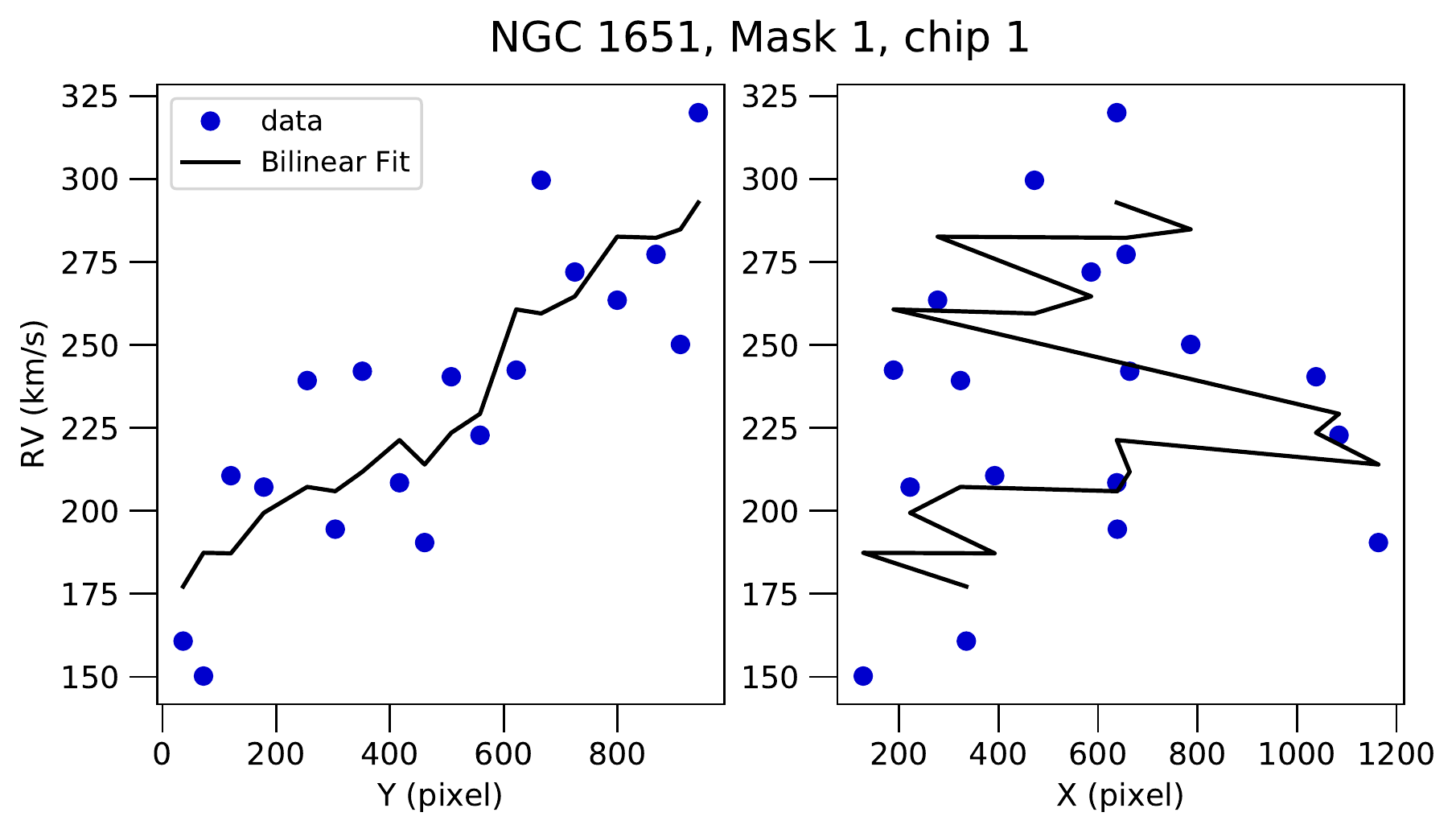}
\caption{Variation of radial velocity of targeted stars as a function of the spatial pixels (Y on the left panel, X on the right panel) on the mask. Here we report the case of NGC 1651, Mask 1 and Chip 1. The black line represents the bilinear fit in both X and Y to the data. See text for more details.} 
\label{fig:flex}
\end{figure}

We noted that the obtained
RVs showed systematic variation along
the Y spatial position of the CCD (while no significant variation on the X position) for each cluster and mask (see also \citealt{pancino10}).
As an example, Figure \ref{fig:flex} shows the
RVs as a function of Y (left panel) and X (right panel) for the Mask 1, chip 1 of NGC 1651. 
Such variation is not a physical property of cluster stars, but it could be 
due to instrument flexures, although we note that the expected instrument flexure is much less according to the FORS2 User Manual\footnote{\url{http://www.eso.org/sci/facilities/paranal/instruments/fors/doc/VLT-MAN-ESO-13100-1543\_P06.pdf}}. 

Hence, to estimate the real RVs of the stars, we corrected for this effect, following the same method reported in \cite{pancino10}.  We applied a bilinear fit to the data in the form of $RV = A + BX + CY$ for each mask and chip of each cluster. This fit is shown in Fig. \ref{fig:flex} as a black line. We calculated the differences between the RVs calculated before and the RVs obtained from the fit and we reported them to each cluster systemic velocity. Systemic velocities were taken from \cite{mucciarelli08}. 
The obtained RVs are reported in Table \ref{tab:info_spectra}. However, given the systematic offsets reported above and the low resolution of the data, we did not use the estimated RVs for the membership selection. 

\bsp	
\label{lastpage}

\clearpage\pagestyle{empty}



\end{document}